\documentclass[final,1p,times]{elsarticle}

\usepackage{graphicx}
\usepackage{amssymb}
\usepackage{natbib}
\usepackage{lineno}
\usepackage{booktabs}
\usepackage{float}
\usepackage{color}
\usepackage{amsmath}
\usepackage{cuted}
\usepackage{lipsum}
\usepackage{subcaption}

\journal{Nucl.\ Instrum.\ Meth.\ Phys.\ Res.\ A}

\begin{document}

\begin{frontmatter}


\title{Transmission Efficiency of the Recoil Mass Spectrometer EMMA at TRIUMF}



\author[label1,label3]{B. Davids}
\ead{davids@triumf.ca}
\author[label1]{N.E. Esker\fnref{label2}}
\author[label5,label1] {J. Jaeyoung}
\author[label5] {Y.K. Kim}
\author[label5,label1]{K. Pak}
\author[label1,label6]{M. Williams\fnref{label4}}


\address[label1]{TRIUMF, 4004 Wesbrook Mall, Vancouver, BC V6T 2A3, Canada}
\fntext[label2]{Present Address: Department of Chemistry, San Jos\'e State University, San Jose, CA, USA}
\address[label3]{Department of Physics, Simon Fraser University, 8888 University Drive, Burnaby, BC V5A 1S6, Canada}
\fntext[label4]{Present Address: Department of Physics, University of Surrey, Guildford GU2 7XH, United Kingdom}
\address[label5]{Department of Nuclear Engineering, Hanyang University, Seoul, Republic of Korea}
\address[label6]{Department of Physics, University of York, Heslington, York YO10 5DD, United Kingdom}

\begin{abstract}

The mean transmission efficiency of the EMMA recoil mass spectrometer at TRIUMF has been measured with 6 different angular apertures at 17 kinetic energy/charge deviations with respect to the central, reference trajectory. Measurements performed using a $^{148}$Gd $\alpha$ source installed at the target position of the spectrometer are compared to ion-optical calculations and Monte Carlo simulations. The transmission efficiency as a function of angle and kinetic energy/charge is described empirically using piecewise Gaussian functions whose parameters are fit to the data.
\end{abstract}


\end{frontmatter}


\section{Introduction}\label{sec:intro}

\begin{figure}
 \includegraphics[width=\textwidth]{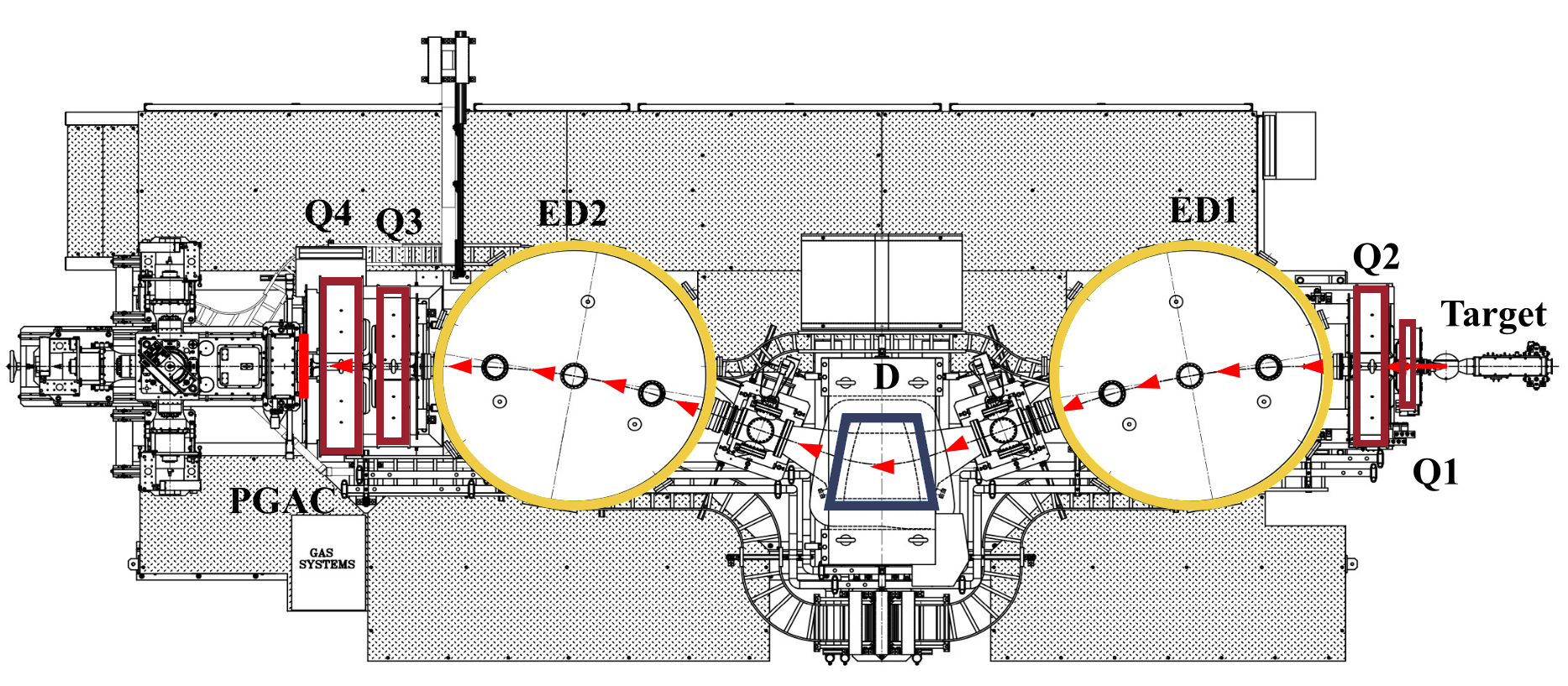}
\caption{Plan view of the Electromagnetic Mass Analyzer EMMA, showing the target chamber, 4 quadrupole magnets (Q1-4), 2 electrostatic deflectors (ED1-2), dipole magnet (D), and the focal plane detector enclosure that contains the parallel grid avalanche counter (PGAC). The arrows indicate the trajectory of an ion moving along the optic axis.} \label{emma_plan}
 \end{figure}

The Electromagnetic Mass Analyzer (EMMA) is a vacuum-mode recoil mass spectrometer designed to separate the products of nuclear reactions from the beam and disperse them in accordance with their mass/charge ($m/q$) ratio at the focal plane. EMMA is situated within the ISAC-II experimental hall at TRIUMF and is designed to be coupled with the TIGRESS $\gamma$-ray spectrometer \cite{hackman14}. It includes four quadrupole magnets, two electrostatic deflectors, and a dipole magnet; a plan view of the spectrometer is shown in Fig.\ \ref{emma_plan}, and details on the components, construction, and commissioning of the spectrometer appear in Ref.\ \cite{davids19}. Some of the experiments that have been and will be performed with EMMA are absolute cross section measurements for nuclear astrophysics \cite{lotay21,williams23,williams25}. They require knowledge of the efficiency for transmitting recoils from the target to the focal plane detectors. Here we present measurements of EMMA's transmission efficiency, which was characterized using an $\alpha$ source positioned at the target location and various apertures installed near the entrance to the spectrometer. The method is described in Section \ref{sec:method} and the experimental procedure is set out in Section \ref{sec:exp}. The measured mean transport efficiencies are presented in Section \ref{sec:results}. Comparisons with ion optical calculations and Monte Carlo simulations are discussed in Section 5 and an empirical description of the transmission efficiency can be found in Section 6. A summary of our findings appears in Section \ref{sec:concl}.

\section{Method}\label{sec:method}

Recoil mass spectrometers have angular and energy acceptances limited by their physical dimensions, beyond which incident charged particles cannot be transmitted to the focal plane detectors unimpeded. The transmission efficiency can be written as a function of the angles with respect to the optic axis and the relative differences in kinetic energy/charge $(\delta T)$ and mass/charge $(\delta m)$ \cite{wollnik87a}. The parameters $\delta T$ and $\delta m$ are defined by

\begin{equation} \label{eqn0}
    \delta T \equiv \frac{T/q - T_{0}/q_{0}}{T_{0}/q_{0}} \quad \mathrm{and} \quad \delta m \equiv \frac{m/q - m_{0}/q_{0}}{m_{0}/q_{0}}
\end{equation}
respectively, where the incident particles have kinetic energy $T$, mass $m$, and charge $q$ and the spectrometer is set for a reference particle with kinetic energy $T_0$, mass $m_0$, and charge $q_0$. Charged particles entering the spectrometer on the optic axis with $T=T_{0}$, $m=m_{0}$ and $q=q_{0}$ continue to travel along the optic axis. The angles with respect to the optic axis in the horizontal and vertical directions are denoted by $\theta$ and $\phi$, respectively. 

As explained in Ref.\ \cite{back96}, the acceptances of a spectrometer need only be determined once, owing to the scaling properties of the trajectories of charged particles in electromagnetic fields. This scaling can be demonstrated by considering a nonrelativistic ion with mass $m$, charge $q$, momentum $\mathbf{p}$, and kinetic energy $T$ moving in a direction perpendicular to spatially separated, homogeneous electric and magnetic fields $\mathbf{E}$ and $\mathbf{B}$. As the ion moves through electric and magnetic fields, its trajectory will bend with radii of curvature given by

\begin{equation} \label{eqn1}
    \rho_{E} = \frac{|\mathbf{p}|v}{q \mathbf{|E|}} \approx \frac{2T}{q\mathbf{|E|}},   
\end{equation}
where $v$ is the speed, and
    
\begin{equation}
    \rho_{B} = \frac{|\mathbf{p}|}{q \mathbf{|B|}} \approx \frac{\sqrt{2mT}}{q \mathbf{|B|}},
\end{equation}
respectively. In order to transport ions with the reference kinetic energy/charge ($T_{0}/q_{0}$) and mass/charge $(m_{0}/q_{0})$ along the optic axis of the spectrometer, the electric and magnetic fields must be tuned such that

\begin{equation}
     \mathbf{|E_{0}|} = \frac{2T_{0}}{q_{0}\rho^{0}_{E}} \quad \mathrm{and} \quad \mathbf{|B_{0}|} = \frac{\sqrt{2m_{0}T_{0}}}{q_{0}\rho^{0}_{B}},
\end{equation}
where $\rho^{0}_{E}$ and $\rho^{0}_{B}$ are defined by the geometry of the electric and magnetic elements. Substituting these field settings into Equation \ref{eqn1} gives

\begin{equation} \label{eqn3}
    \rho_{E} = \frac{T/q}{T_{0}/q_{0}} \rho^{0}_{E} \quad \mathrm{and} \quad  \rho_{B} =  \sqrt{ \frac{m/q}{m_{0}/q_{0}} \frac{T/q}{T_{0}/q_{0}}}\rho^{0}_{B}.
\end{equation}

Therefore, the trajectories of ions in the spectrometer depend only on their $m/q$ and $T/q$ ratios with respect to ions that traverse the spectrometer on the reference trajectory along the optic axis. Similar arguments apply to ions on off-axis trajectories. The transport efficiency, $\epsilon$, is then a function dependent only on the angles $\theta$ and $\phi$ and the ratios $(T/q)/(T_{0}/q_{0})$ and $(m/q)/(m_{0}/q_{0})$, which can be more conveniently expressed via Equation \ref{eqn0} in terms of $\delta T$ and $\delta m$. That is,

\begin{equation} \label{eqn4}
    \epsilon \Bigg( \frac{T/q}{T_{0}/q_{0}}, \frac{m/q}{m_{0}/q_{0}}, \theta, \phi \Bigg) \quad \mathrm{or} \quad \epsilon( \delta T, \delta m, \theta, \phi).
\end{equation}

\section{Measurement Procedure} \label{sec:exp}

In this section we describe how the transmission efficiency of EMMA was determined as a function of each parameter in Equation \ref{eqn4}. Since it is relatively trivial to change the field settings of EMMA, we varied $\delta T$ and $\delta m$ by changing the parameters $T_{0}$ and $m_{0}$ for a single charge state setting $q_{0}=q$. As absolute cross section measurements with EMMA invariably utilize a single mass and charge state, we determined the transmission efficiency as a function of $\delta T$, $\theta$ and $\phi$, while leaving the parameter $\delta m = 0$. Ideally one would like to find the transmission efficiency $\epsilon(\delta T,\theta, \phi)$ for all values of ($\theta$, $\phi$) and any value of $\delta T$ within the acceptance. However, measuring $\epsilon$ at a very fine, granular level is time-consuming and impractical. Therefore we carried out mean transmission efficiency measurements at $\delta T$ settings from -0.20 to 0.50 in increments of 0.025, 0.05, and 0.10.

\begin{figure}
 \includegraphics[width=\textwidth]{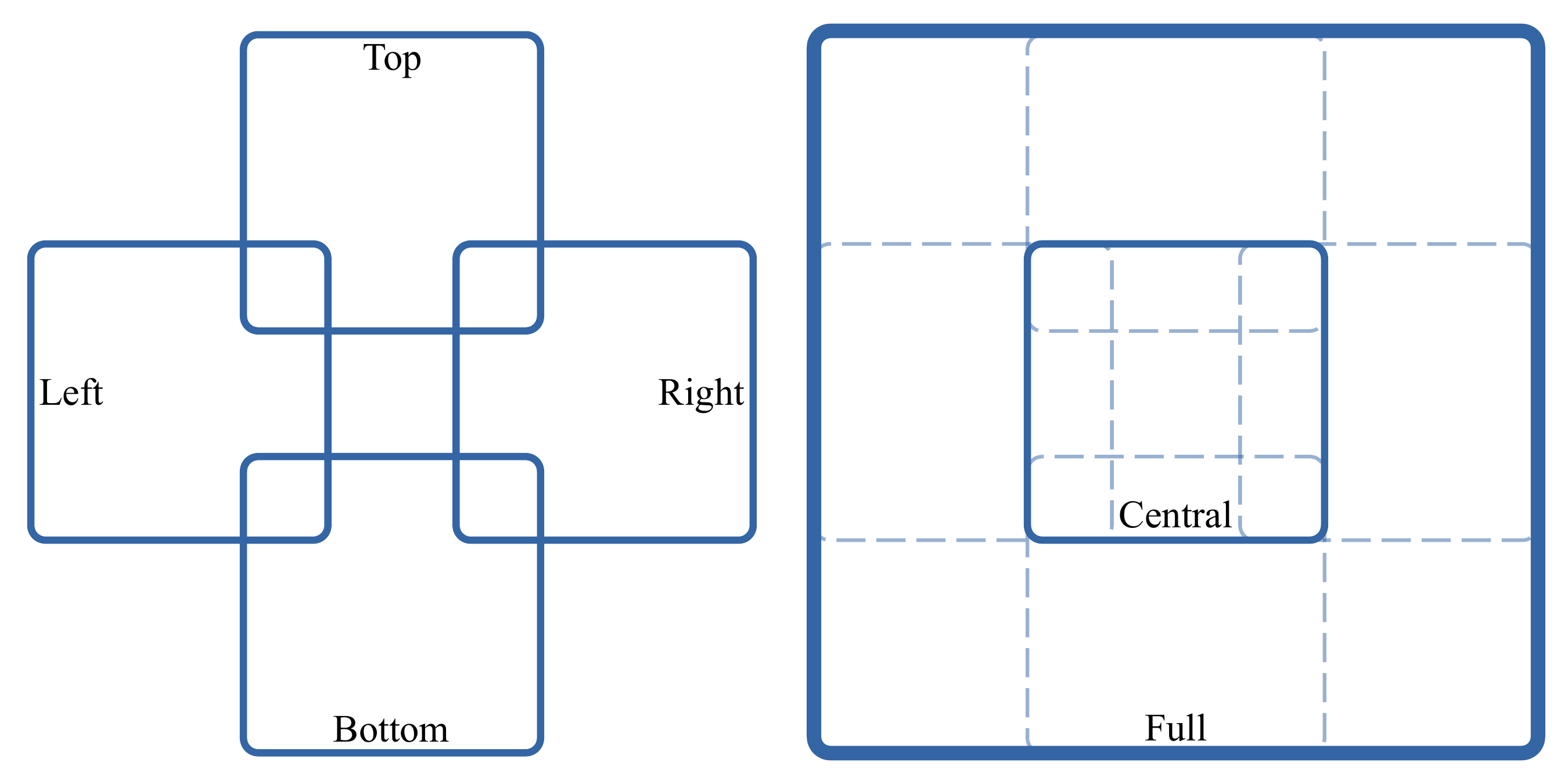}
\caption{Schematic rendering of the angular apertures used to measure the transmission efficiency of EMMA, indicating their orientations and relative sizes. The 4 off-axis apertures are shown on the left, while the 2 on-axis apertures appear on the right.}
\label{apertures}
 \end{figure}

For these measurements we utilized a 1.18(1) kBq $^{148}$Gd $\alpha$ source ($T_{1/2}=71.1\pm1.2$~y) mounted at the target position. This radionuclide decays 100\% of the time via 3.183 MeV $\alpha$ particles \cite{akovali98, nica14}. Employing a radioactive source rather than elastically scattered beam or target ions is advantageous for two main reasons. First, all of the $\alpha$ particles emitted by the source have the same charge of $2e$, whereas the charge state distributions of elastically scattered beam and target ions are not precisely known, thereby requiring additional measurements. Second, source measurements do not require an accelerator and can therefore be performed over the extended times required for large $|\delta T|$ settings without preventing accelerated beam experiments at other facilities. The entrance angles of the $\alpha$ particles were defined by apertures fixed within the target chamber 80~mm downstream of the source. The measured aperture dimensions and the horizontal and vertical angles $\theta$ and $\phi$ subtended by each of the 6 apertures are specified in Table \ref{table:apertures}, along with their solid angles. There is one large aperture, labelled ``Full", which encompasses the entire angular range subtended by each of five smaller apertures, labelled as: ``Central", ``Left", ``Right", ``Top" and ``Bottom"; the last four apertures are centred off-axis. The orientations and relative sizes of the apertures are depicted in Fig.\ \ref{apertures}.

\begin{table}[h!]
\centering
\begin{tabular}{c c c c c c c c}
	\toprule[1pt]\midrule[0.3pt]
	\noalign{\smallskip}
	Aperture & Width & Height & $\theta_{\textrm{min}}$ & $\theta_{\textrm{max}}$ & $\phi_{\textrm{min}}$ & $\phi_{\textrm{max}}$ & Solid Angle\\ \midrule
	Full & 8.31 & 8.29 & -2.99 & 2.97 & -2.97 & 2.97 & 10.8(4)\\
	Central & 3.41 & 3.40 & -1.29 & 1.16	& -1.25 & 1.19 & 1.82(18) \\
	Left & 3.38 & 3.37 & -3.02 & -0.59 & -1.21 & 1.20 & 1.79(18)\\
	Right	 & 3.37 & 3.36 & 0.60 & 3.02 & -1.19 & 1.22 & 1.78(18)\\
	Bottom & 3.37 & 3.37 & -1.21 & 1.21 & -3.00 & -0.59 & 1.78(18)\\
	Top & 3.37 & 3.35 & -1.22 & 1.20 & 0.60	 & 3.00 & 1.77(18)\\
	\noalign{\smallskip} 
	\midrule[0.3pt]\bottomrule[1pt]
\end{tabular}
\caption{Measurements of the dimensions in mm, horizontal ($\theta$) and vertical ($\phi$) angles with respect to the optic axis in $^{\circ}$, and the subtended solid angles in msr of the various angular apertures used in the transmission efficiency measurements. The systematic uncertainty in each dimension is 0.24~mm and that in each plane angle is 0.12$^\circ$.}
\label{table:apertures}
\end{table}

Systematic uncertainties arise from uncertainties in the aperture dimensions and the aperture location, due to the small mechanical tolerance in the positioning of screws used to fix the apertures in place. The solid angle subtended by aperture $i$, using an appropriate small angle approximation, is given by
\begin{equation}
\Omega_{\mathrm{i}} \approx \frac{\mathrm{w_i \times h_i}}{\mathrm{D}^2},
\label{eq:solidangle}
\end{equation}
where w$_\mathrm{i}$ and h$_\mathrm{i}$ are the width and height of aperture $i$ and $D = 79.85(35)$~mm is the distance from the $\alpha$ source to the aperture.

The relative systematic uncertainty of the solid angles is given by the equation

\begin{equation}
\left(\frac{\delta \Omega_{\mathrm{i}}}{\Omega_{\mathrm{i}}}\right)^2=\left(\frac{\delta \mathrm{w_i}}{\mathrm{w_i}}\right)^2+\left(\frac{\delta \mathrm{h_i}}{\mathrm{h_i}}\right)^2+4 \Omega_i^2 \left(\frac{\delta \mathrm{D}}{\mathrm{D}}\right)^2,
\end{equation}
where $\delta \mathrm{w_i}$=$\delta \mathrm{h_i} = 0.24$~mm, stemming from the roughness of the aperture edges, and $\delta D=0.35$~mm accounts for the play in the aperture screws. The relative systematic uncertainties of the solid angles of the Central and off-axis apertures are 10\%, while that of the Full aperture is 4\% due to its larger size.

The $\alpha$ particles transmitted through the spectrometer were detected by a position-sensitive parallel grid avalanche counter (PGAC) and stopped in a 3000 mm$^{2}$ ion-implanted silicon detector. For technical details regarding the focal plane detectors of EMMA, the reader is referred to Ref.\ \cite{davids19}. The measured mean transmission efficiency through each angular aperture $\langle\epsilon \rangle_i$ is defined as the number of $\alpha$ particles reaching the focal plane, $N_{\mathrm{fp},i}$ divided by the number of $\alpha$ particles entering the spectrometer through each aperture, $N_{\alpha,i}$,
\begin{equation}
    \langle\epsilon \rangle_i \equiv \frac{N_{\mathrm{fp},i}}{N_{\alpha,i}}.
\end{equation}
For each aperture $i$, the number of $\alpha$ particles reaching the focal plane is related to the number of detected $\alpha$ particles $N_{det,i}$ by

\begin{equation}
    N_{det,i}=\eta N_{\mathrm{fp},i},
\end{equation}
where the intrinsic detector efficiency $\eta=0.936$ accounts for the $93.6(3)\%$ transparency of the PGAC wire grid. The total number of $\alpha$ particles entering the spectrometer through aperture $i$ is given by:

\begin{equation}
 N_{\alpha,i} = \frac{A t \Omega_i}{4 \pi},
\label{eq:meas}
\end{equation}
where $A$ is the activity of the source, $t$ is the data collection time, and $\Omega_i$ is the solid angle subtended by the aperture. The data collection time was varied in order to gather sufficient statistics for each $\delta T$ setting and aperture. Due to the low event rates, dead time was negligible in these measurements.

The procedure of acquiring data started with the installation of an angular aperture and included a series of measurements at different $\delta T$ settings. The process was repeated for each of the 6 angular apertures. For monoenergetic ions of a single mass and charge, the mean transmission efficiency can also be written as the angle-dependent transmission efficiency averaged over the angles subtended by the aperture $i$, 
\begin{equation}
 \langle\epsilon \rangle_i = \frac{\iint_{\omega_{i}} \epsilon(\theta, \phi)\mathrm{d}\Omega}{\iint_{\omega_{i}} \mathrm{d}\Omega},
 \label{eq:average}
\end{equation}
where $\omega_{i}$ and $\epsilon(\theta,\phi)$ schematically denote the angular limits of aperture $i$ and $\epsilon(\delta T, \delta m = 0, \theta,\phi)$ evaluated at  the specified $\delta T$ value, respectively.

\section{Experimental Results} \label{sec:results}

Mean transmission efficiency measurements for each angular aperture at various $\delta T$ settings are shown in Table \ref{tab:full} along with their statistical uncertainties. The relative systematic uncertainties in both the source activity (0.8\%) and the PGAC transparency (0.3\%) are common to all measurements. They are added in quadrature with the systematic uncertainty due to the solid angle of the aperture and the statistical uncertainty to determine the total uncertainty of each mean transmission efficiency measurement. The uncertainty in the solid angle of an angular aperture is common to all measurements with that aperture but not to all measurements at a given $\delta T$ setting.

For all but a few data sets at the most extreme values of $\delta T$, at least one thousand events were recorded, resulting in a statistical uncertainty not exceeding 3$\%$. Therefore, the uncertainties in the transmission efficiencies for the small apertures are dominated by systematic uncertainties, whereas the Full aperture uncertainties have roughly equal contributions from systematic and statistical sources.

\begin{table*}
\begin{center}
\begin{tabular}{l l l l l l l}
\hline
\multicolumn{1}{c}{$\delta T$} & \multicolumn{1}{c}{Full} & \multicolumn{1}{c}{Central} & \multicolumn{1}{c}{Left} & \multicolumn{1}{c}{Right} & \multicolumn{1}{c}{Bottom} & \multicolumn{1}{c}{Top}\\
\hline
-0.20 & 0.046(1) & 0.073(3) & 0.025(2) & 0.145(4) & 0.176(4) & 0.011(1)\\
-0.15 & 0.314(10) & 0.435(7) & 0.285(7) & 0.608(13) & 0.573(17) & 0.132(4)\\
-0.10 & 0.608(17) & 0.810(21) & 0.573(18)& 0.879(26) & 0.745(21) & 0.453(13)\\
-0.075 & 0.661(10) & & & & &\\
-0.05 & 0.706(14) & 0.901(20) & 0.726(21) & 0.926(26) & 0.772(23) & 0.593(18)\\
-0.025 & 0.710(14) & & & & &\\
0.0 & 0.691(9) & 0.908(13) & 0.831(8) & 0.775(8) & 0.651(19) & 0.583(16)\\
0.025 & 0.673(13) & & & & &\\
0.05 & 0.627(6) & 0.834(24) & 0.839(25) & 0.569(18) & 0.520(14) & 0.548(17)\\
0.10 & 0.571(14) & 0.708(18) & 0.853(26) & 0.405(13) & 0.396(12)& 0.502(15)\\
0.15 & 0.472(7) & 0.611(18) & 0.804(25) & 0.278(8) & 0.270(8) & 0.407(12)\\
0.20 & 0.366(8) & 0.447(13) & 0.749(22) & 0.143(4) & 0.160(4) & 0.331(6)\\
0.25 & 0.262(6) & 0.327(6) & 0.614(19) & & 0.073(2) &\\
0.30 & 0.192(3) & 0.215(6) & 0.499(13) & & 0.036(2)& 0.140(4)\\
0.35 & 0.131(2) & & 0.367(7) & 0.0014(5) & &\\
0.40 & 0.091(2) & & & & &\\
0.50 & 0.036(1) & & & & &\\
\\
 
\hline

\end{tabular}
\caption{\label{tab:full}Measured mean efficiencies for the transmission of $\alpha$ particles through 6 angular apertures at various relative kinetic energy/charge deviation ($\delta T$) settings presented with their statistical uncertainties. Apart from a few settings with small transmission efficiencies, relative statistical errors ranged from 1-3\%. Systematic uncertainties are discussed in Sections \ref{sec:exp} and \ref{sec:results}.}
\end{center}
\end{table*}

\section{Comparison with Simulations}

\begin{figure}
\begin{subfigure}{0.5\textwidth}
 \includegraphics[width=\linewidth, height=0.588\linewidth]{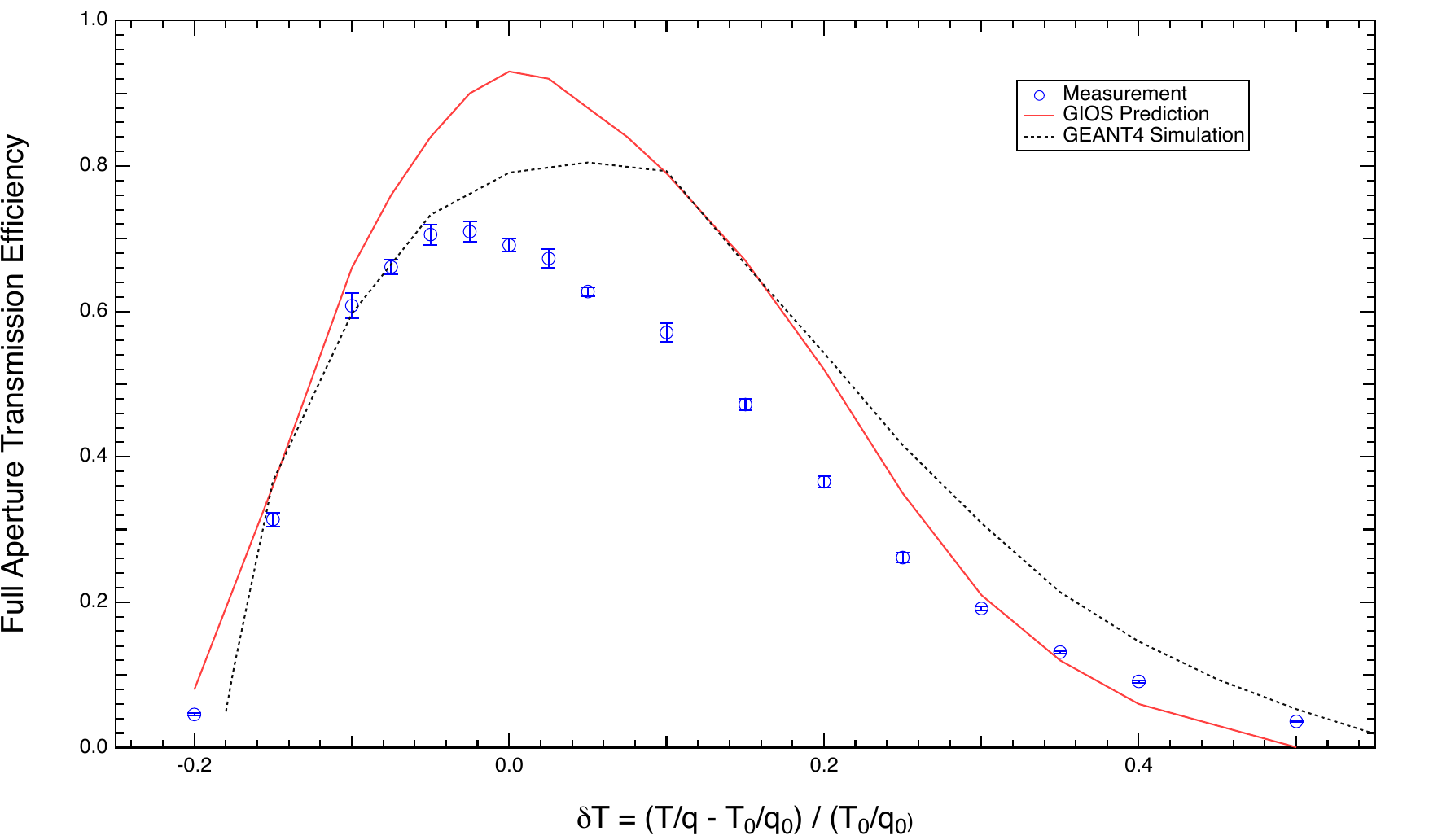}
\caption{Full Aperture} \label{fig:acceptance_full}
 \end{subfigure}
\begin{subfigure}{0.5\textwidth}
  \includegraphics[width=\linewidth, height=0.588\linewidth]{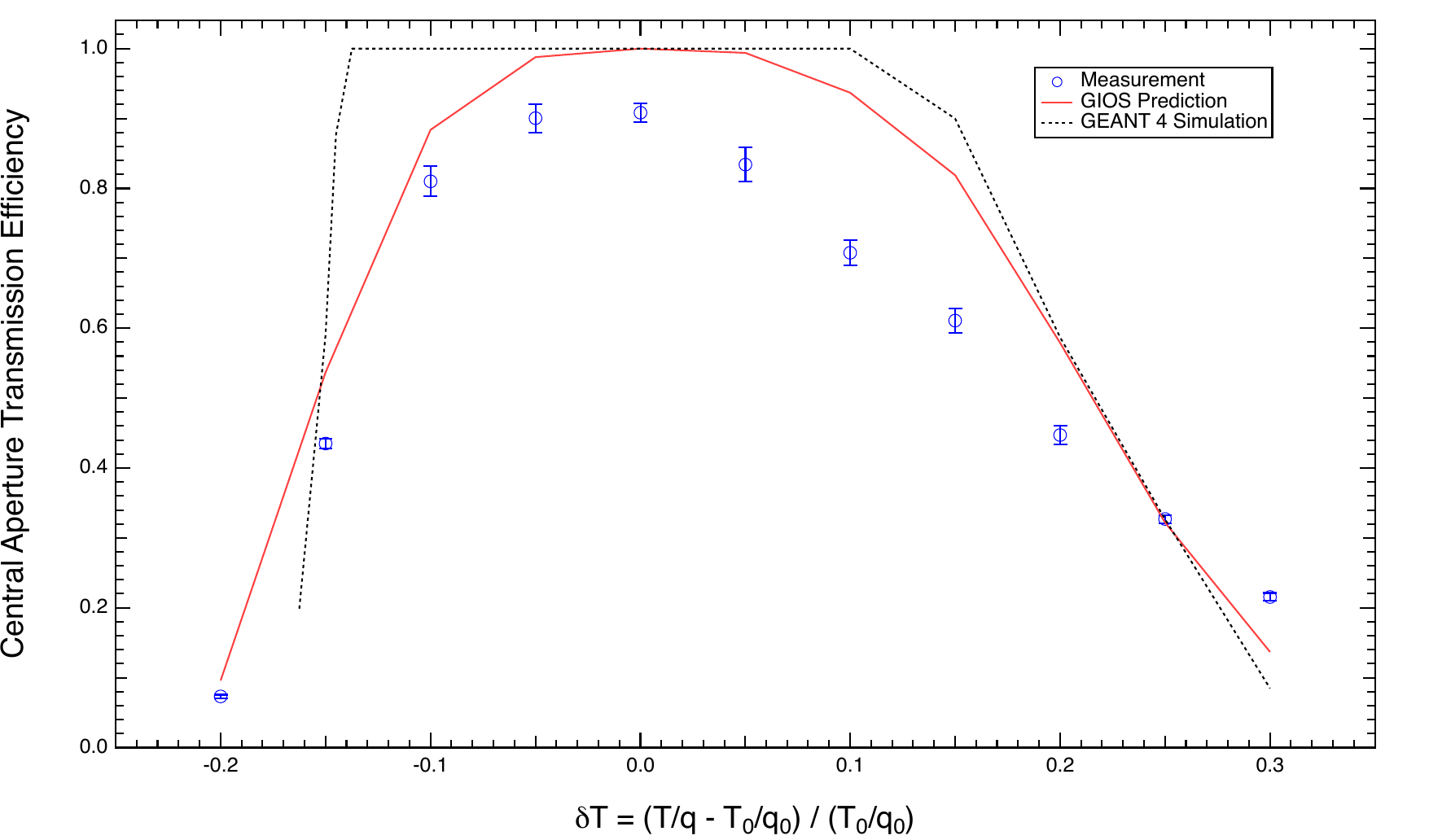}
\caption{Central Aperture} \label{fig:acceptance_central}
 \end{subfigure}
 \caption{Measured mean transmission efficiency as a function of $\delta T$ with statistical errors. The solid curve represents the predicted transmission efficiency calculated with GIOS. The dotted curve shows the transmission efficiency calculated with a GEANT4 simulation.}
 \label{fig:acceptance_full_central}
 \end{figure}

\begin{figure}
\begin{subfigure}{0.5\textwidth}
 \includegraphics[width=\linewidth, height=0.588\linewidth]{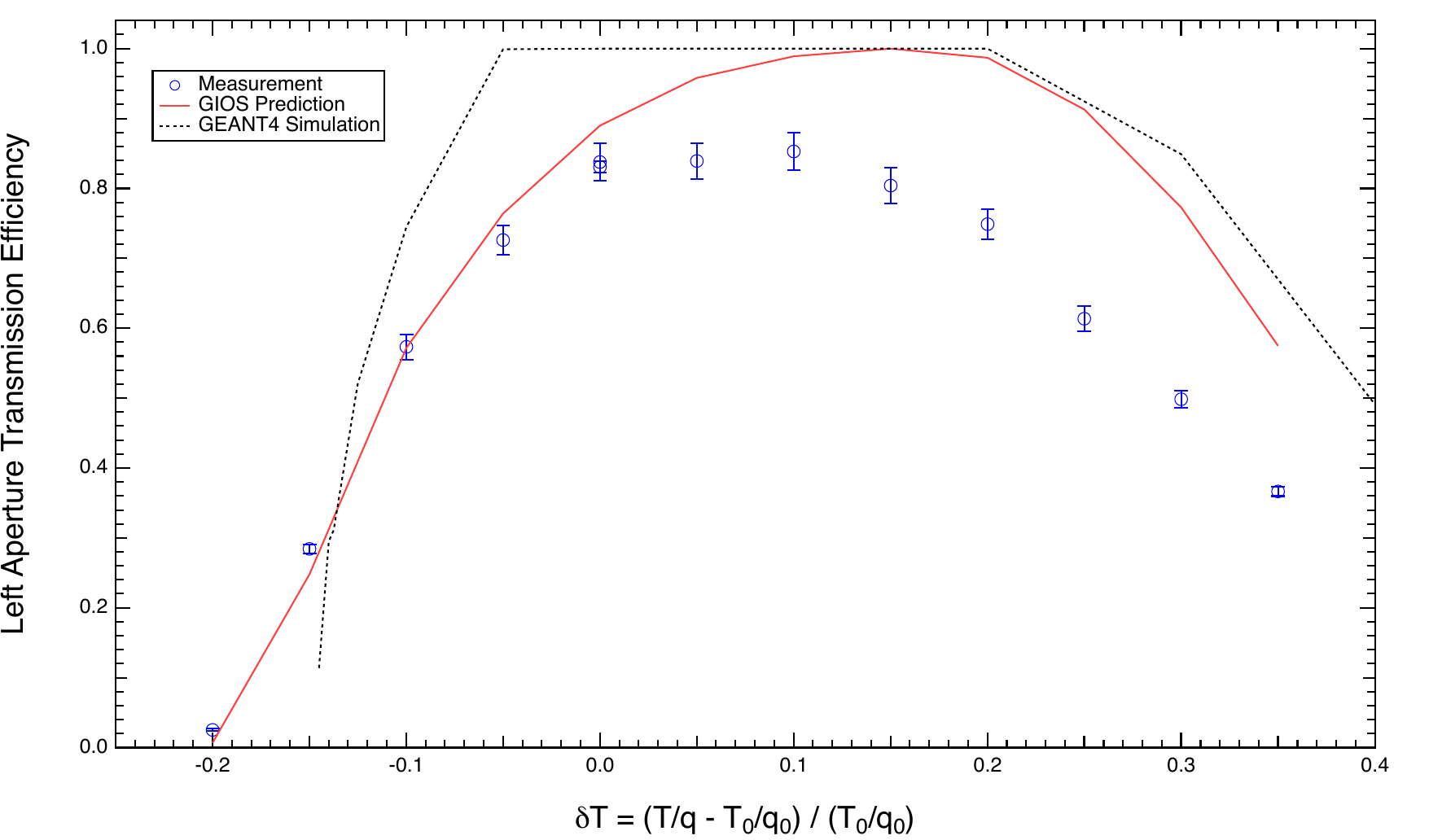}
\caption{Left Aperture} \label{fig:acceptance_left}
 \end{subfigure}
\begin{subfigure}{0.5\textwidth}
  \includegraphics[width=\linewidth, height=0.588\linewidth]{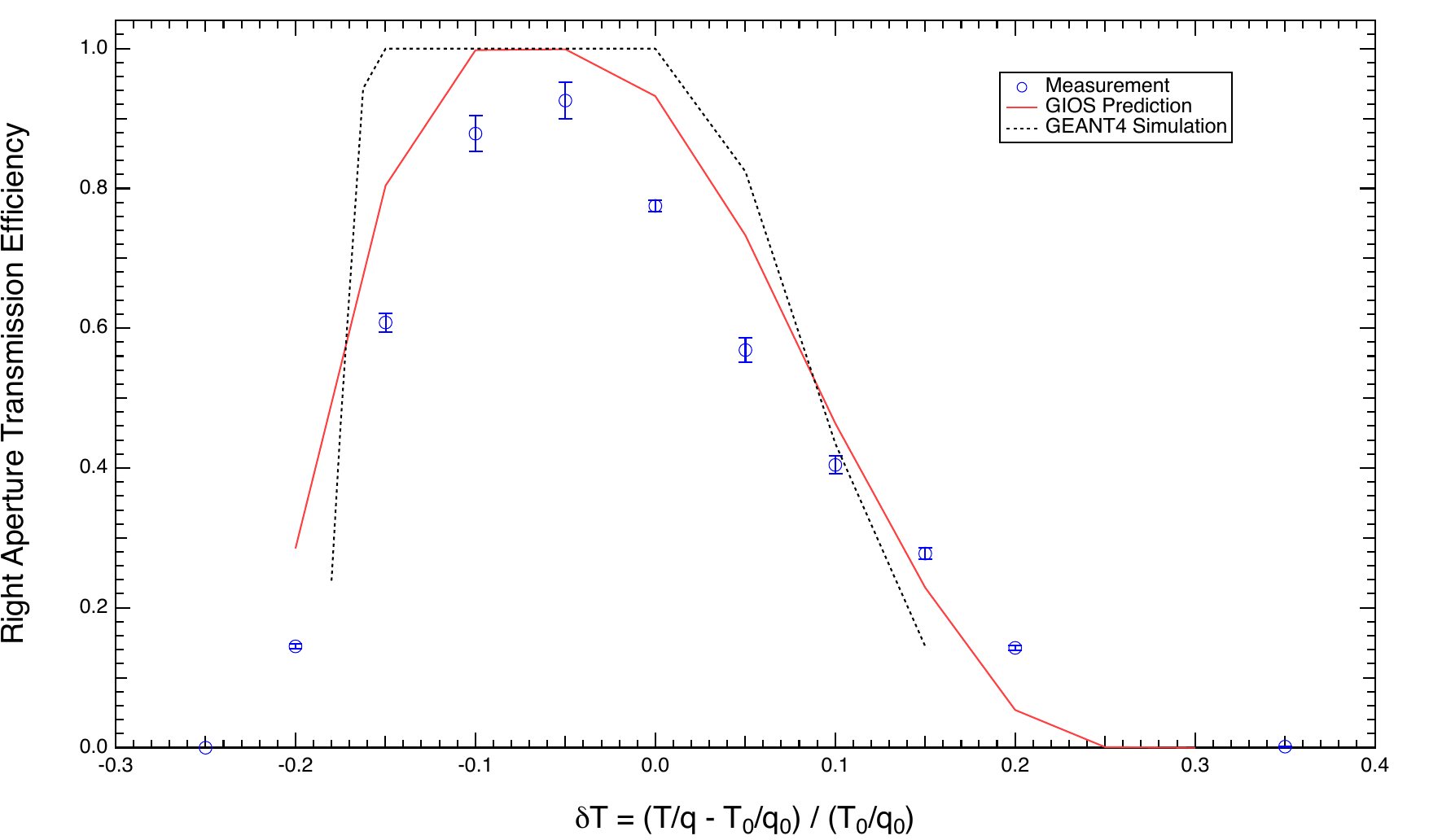}
\caption{Right Aperture} \label{fig:acceptance_right}
 \end{subfigure}
 \caption{Measured mean transmission efficiency as a function of $\delta T$ with statistical errors. The solid curve represents the predicted transmission efficiency calculated with GIOS. The dotted curve shows the transmission efficiency calculated with a GEANT4 simulation.}
  \label{fig:acceptance_left_right}
 \end{figure}

\begin{figure}
\begin{subfigure}{0.5\textwidth}
 \includegraphics[width=\linewidth, height=0.588\linewidth]{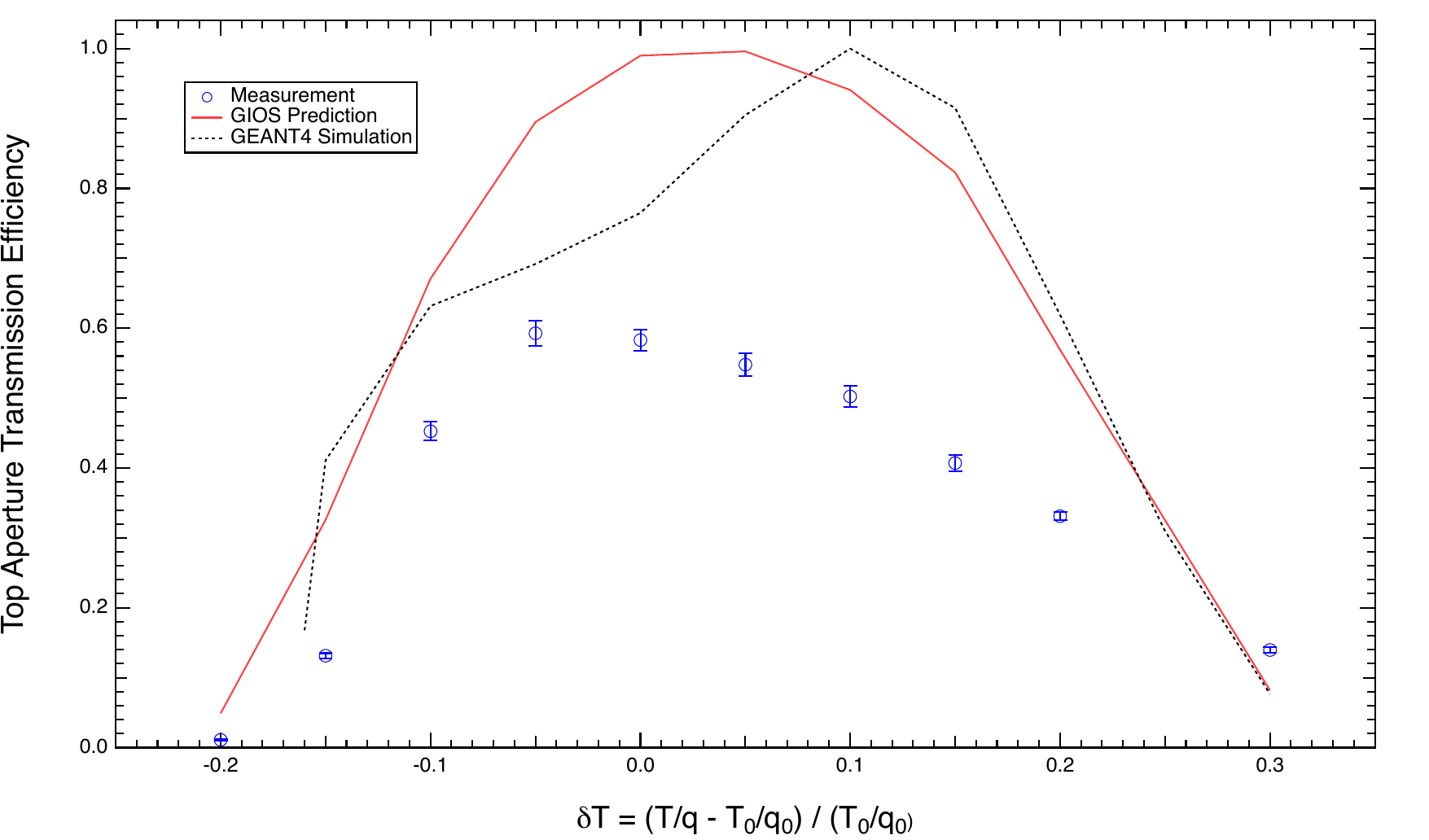}
\caption{Top Aperture} \label{fig:acceptance_top}
 \end{subfigure}
\begin{subfigure}{0.5\textwidth}
  \includegraphics[width=\linewidth, height=0.588\linewidth]{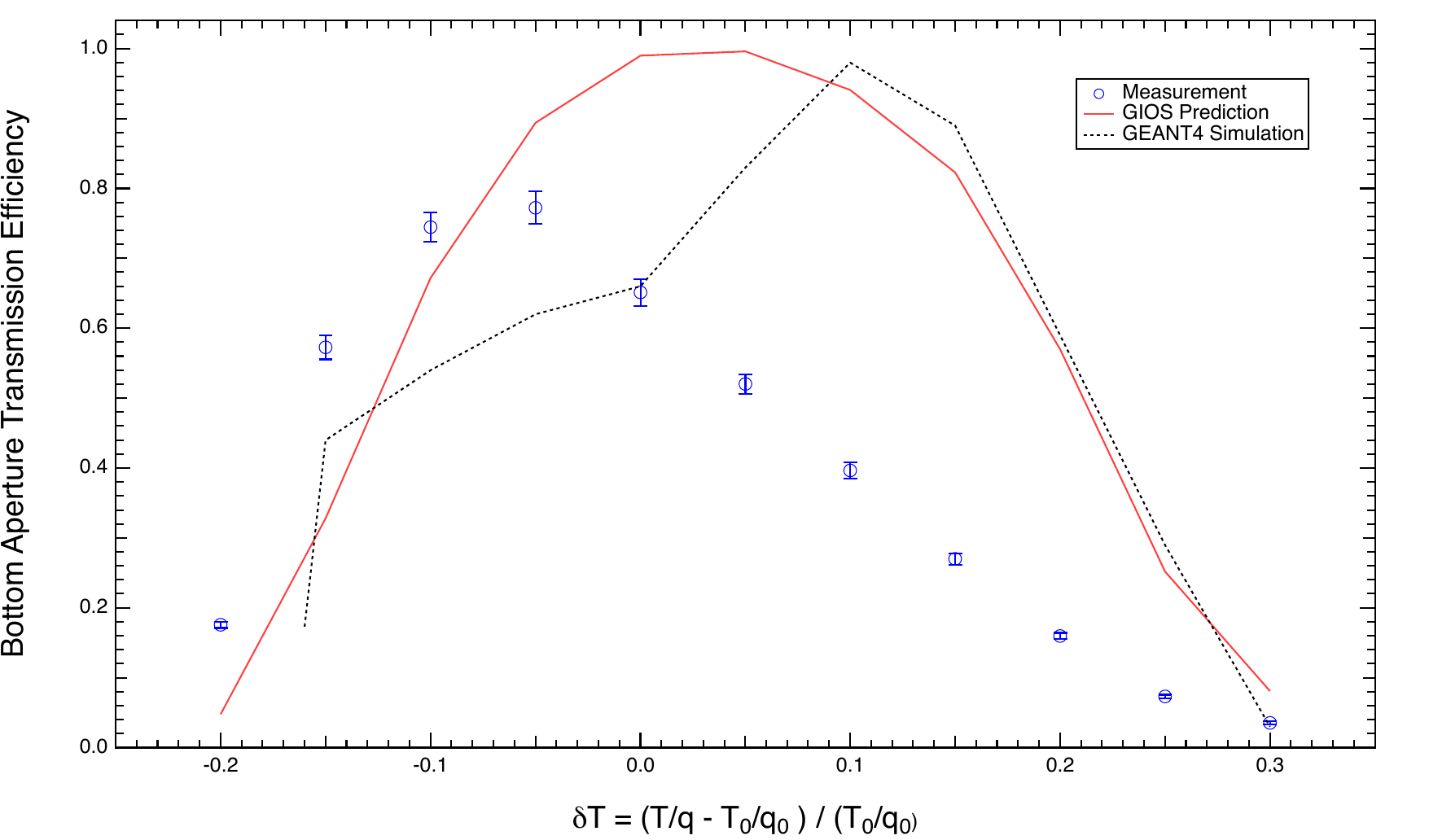}
\caption{Bottom Aperture} \label{fig:acceptance_bottom}
 \end{subfigure}
 \caption{Measured mean transmission efficiency as a function of $\delta T$ with statistical errors. The solid curve represents the predicted transmission efficiency calculated with GIOS. The dotted curve shows the transmission efficiency calculated with a GEANT4 simulation.}
   \label{fig:acceptance_top_bottom}
 \end{figure}

A model of the spectrometer was constructed using the GEANT4 simulation toolkit \cite{geant4_ref}, incorporating all the electromagnetic elements and vacuum components. The idealized fields produced by the ion-optical components of EMMA \cite{davids05} were implemented with overlapping fields because of the minimal distances between the effective field boundaries of the quadrupole magnets in each doublet. The measured mean transmission efficiency through each aperture as a function of $\delta T$ is shown in Figures \ref{fig:acceptance_full_central}, \ref{fig:acceptance_left_right}, and \ref{fig:acceptance_top_bottom}. For comparison, the  transmission efficiency calculated using the ion-optics code GIOS \cite{wollnik87} employed to design the spectrometer, based on the as-built dimensions of the components \cite{davids19} is shown along with the results of the GEANT4 simulations.

It appears from Figures \ref{fig:acceptance_full}, \ref{fig:acceptance_top}, and \ref{fig:acceptance_bottom} that the relative differences between the measurements and the calculations are largest for $\delta T \geq 0$, where the transmission efficiency falls off more rapidly than expected as a function of vertical angle $(\phi)$. The GEANT4 simulations exhibit reduced transmission at $\delta T=0$ through the Full, Top, and Bottom apertures compared to the GIOS prediction. These losses seem to be due to trajectories intersecting the top and bottom of the vacuum chamber of the third quadrupole magnet (Q3). Inspection of the tracks through the simulation suggest that the vertical cross-over occurs upstream of the centre of the dipole magnet, resulting in a larger vertical extent of the transmitted ions within the Q3 vacuum chamber in the GEANT4 simulations relative to the GIOS predictions. With the exception of these calculated losses in Q3 and a sharper transmission cutoff through the Left and Right apertures at $\delta T<0$, the GEANT4 simulations and the GIOS predictions agree tolerably.

\section{Empirical Description}

Given the significant differences between measured and predicted transmission efficiencies as a function of $\delta T$, $\theta$ and $\phi$, we cannot use GIOS predictions or GEANT4 simulations to accurately obtain the transmission efficiency for most experiments. This is likely a consequence of the manufacturing defects of the electrostatic deflectors described in Ref.\ \cite{davids19}. Instead, we adopt an empirical approach and estimate the transmission efficiency at 9 different $\delta T$ values between -0.2 and 0.2 as functions of $\theta$ and $\phi$ using piecewise Gaussian functions whose parameters are adjusted to reproduce the measurements. The transmission efficiency as a function of $\delta T$ between different $\delta T$ values at which there are empirical models is obtained by linearly interpolating between the empirical models at the nearest $\delta T$ settings.
  
 After investigating many possible empirical fitting functions, piecewise two dimensional Gaussian functions were found to best fit the transmission measurements for the given number of free parameters while meeting the physical requirement that the transmission efficiency $\epsilon(\theta,\phi,\delta T)\in[0,1]$. Three types of empirical Gaussian fitting functions were studied: 4 parameter piecewise Gaussians centred on $(\theta,\phi)=(0,0)$, $\epsilon_{pg}(\theta,\phi)$; 4 parameter horizontally offset Gaussians centred on $(\theta,\phi)=(\theta_0,0)$, $\epsilon_{og}(\theta,\phi)$; and 6 parameter, arbitrarily normalized, horizontally offset Gaussians also centred on $(\theta,\phi)=(\theta_0,0)$, $\epsilon_{ng}(\theta,\phi)$. In the 4 parameter piecewise Gaussian fits, the standard deviations $\sigma_L$, $\sigma_R$, $\sigma_B$, and $\sigma_T$ were varied, while in the horizontally offset piecewise Gaussian fits, the standard deviations $\sigma_R$, $\sigma_B$, and $\sigma_T$ were varied along with the horizontal offset $\theta_0$. The standard deviation $\sigma_\mathrm{L}$ of the offset piecewise Gaussian function was fixed at $10^\circ$, an arbitrarily large value, in order to maintain 4 fit parameters. In the 6 parameter fits, the 4 standard deviations were varied along with the horizontal offset and an overall normalization $N$. The explicit form of the two dimensional piecewise Gaussian function appears in Equation~\ref{eq:pgaussian}, while that of the offset Gaussian is in Equation~\ref{eq:ogaussian}, and the normalized offset Gaussian appears in Eq.\ ~\ref{eq:ngaussian}.

\begin{equation}
\epsilon_{pg}(\theta,\phi)= 
\begin{cases}
\exp{[-\theta^2/(2\sigma^2_\mathrm{R})]} \exp{-[\phi^2/(2\sigma^2_\mathrm{T})]},& \text{if } \theta \geq 0 \; \mathrm{and} \; \phi \geq 0 \\
\exp{[-\theta^2/(2\sigma^2_\mathrm{R})]} \exp{-[\phi^2/(2\sigma^2_\mathrm{B})]},& \text{if } \theta \geq 0 \; \mathrm{and} \; \phi < 0 \\
\exp{[-\theta^2/(2\sigma^2_\mathrm{L})]} \exp{-[\phi^2/(2\sigma^2_\mathrm{B})]},& \text{if } \theta < 0 \; \mathrm{and} \; \phi < 0 \\
\exp{[-\theta^2/(2\sigma^2_\mathrm{L})]} \exp{-[\phi^2/(2\sigma^2_\mathrm{T})]},& \text{if } \theta < 0 \; \mathrm{and} \; \phi \geq 0. \\
\end{cases}
\label{eq:pgaussian}
\end{equation}

\begin{equation}
\epsilon_{og}(\theta,\phi)= 
\begin{cases}
\exp{[-(\theta-\theta_0)^2/(2\sigma^2_\mathrm{R})]} \exp{-[\phi^2/(2\sigma^2_\mathrm{T})]},& \text{if } \theta \geq \theta_0 \; \mathrm{and} \; \phi \geq 0 \\
\exp{[-(\theta-\theta_0)^2/(2\sigma^2_\mathrm{R})]} \exp{-[\phi^2/(2\sigma^2_\mathrm{B})]},& \text{if } \theta \geq \theta_0 \; \mathrm{and} \; \phi < 0 \\
\exp{[-(\theta-\theta_0)^2/(2(10^\circ)^2)]} \exp{-[\phi^2/(2\sigma^2_\mathrm{B})]},& \text{if } \theta < \theta_0 \; \mathrm{and} \; \phi < 0 \\
\exp{[-(\theta-\theta_0)^2/(2(10^\circ)^2)]} \exp{-[\phi^2/(2\sigma^2_\mathrm{T})]},& \text{if } \theta < \theta_0 \; \mathrm{and} \; \phi \geq 0. \\
\end{cases}
\label{eq:ogaussian}
\end{equation}

\begin{equation}
\epsilon_{ng}(\theta,\phi)= 
\begin{cases}
N \exp{[-(\theta-\theta_0)^2/(2\sigma^2_\mathrm{R})]} \exp{-[\phi^2/(2\sigma^2_\mathrm{T})]},& \text{if } \theta \geq \theta_0 \; \mathrm{and} \; \phi \geq 0 \\
N \exp{[-(\theta-\theta_0)^2/(2\sigma^2_\mathrm{R})]} \exp{-[\phi^2/(2\sigma^2_\mathrm{B})]},& \text{if } \theta \geq \theta_0 \; \mathrm{and} \; \phi < 0 \\
N \exp{[-(\theta-\theta_0)^2/(2\sigma^2_\mathrm{L})]} \exp{-[\phi^2/(2\sigma^2_\mathrm{B})]},& \text{if } \theta < \theta_0 \; \mathrm{and} \; \phi < 0 \\
N \exp{[-(\theta-\theta_0)^2/(2\sigma^2_\mathrm{L})]} \exp{-[\phi^2/(2\sigma^2_\mathrm{T})]},& \text{if } \theta < \theta_0 \; \mathrm{and} \; \phi \geq 0. \\
\end{cases}
\label{eq:ngaussian}
\end{equation}

The empirical Gaussian models were optimized by integrating the transmission efficiency of each model over the 6 angular apertures according to Equation \ref{eq:average} and computing the $\chi^2$ value with respect to the measured mean transmission efficiencies defined in Eq.\ \ref{eq:meas}. Initial fits were carried out via comprehensive grid searches throughout the parameter spaces of each model. Parameters that globally minimize $\chi^2$ were determined via least squares fitting with the MIGRAD algorithm of the CERN Library minimization routine MINUIT \cite{james75}.

\begin{figure}
 \includegraphics[width=0.75\textwidth]{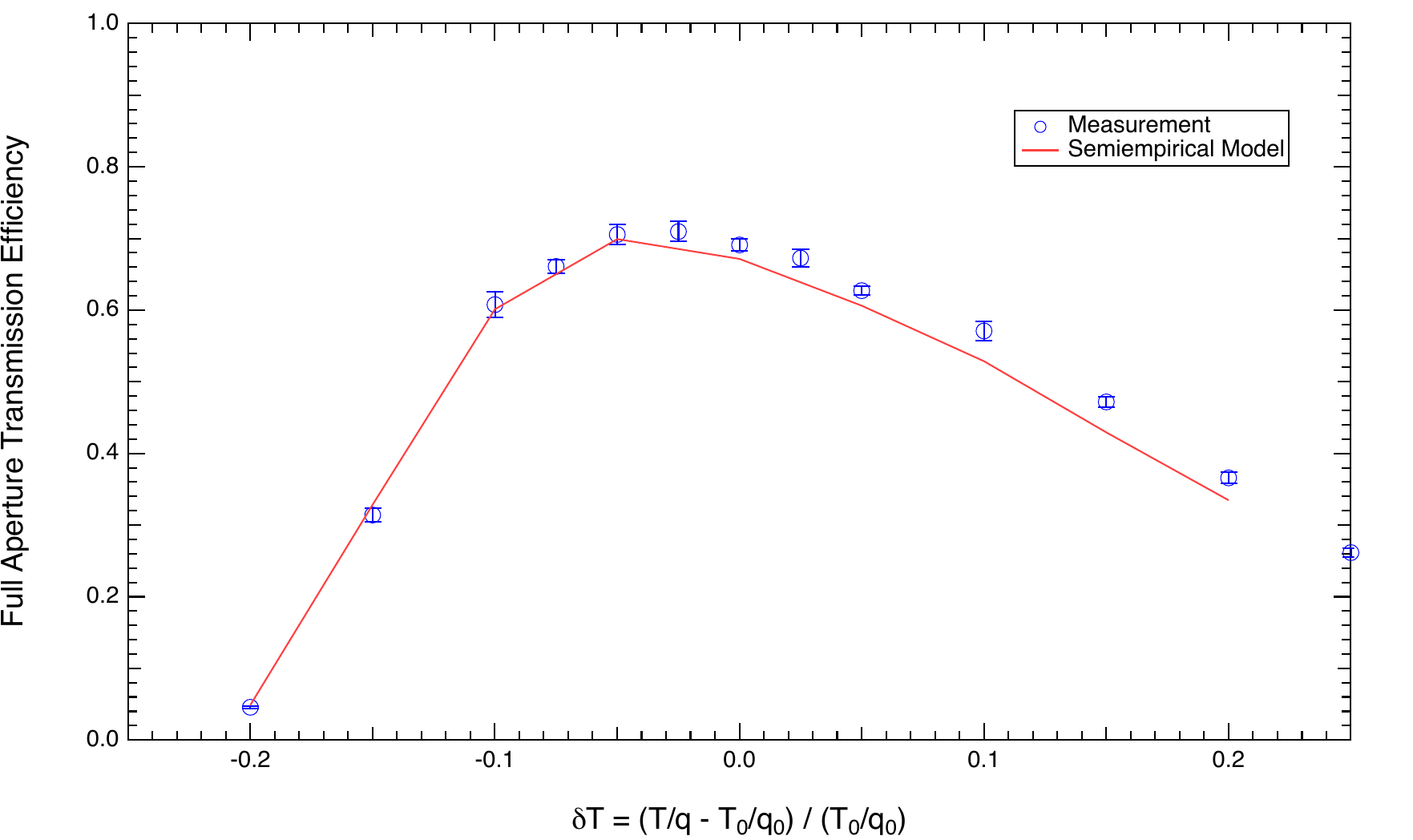}
\caption{Measured mean transmission efficiency as a function of $\delta T$ for the Full angular aperture described in Table \ref{table:apertures} with statistical errors. The solid curve shows the transmission efficiency calculated with the best-fitting empirical model calculations at 9 measured $\delta T$ settings.} \label{fig:full_semi}
 \end{figure}

\begin{figure}
 \includegraphics[width=0.75\textwidth]{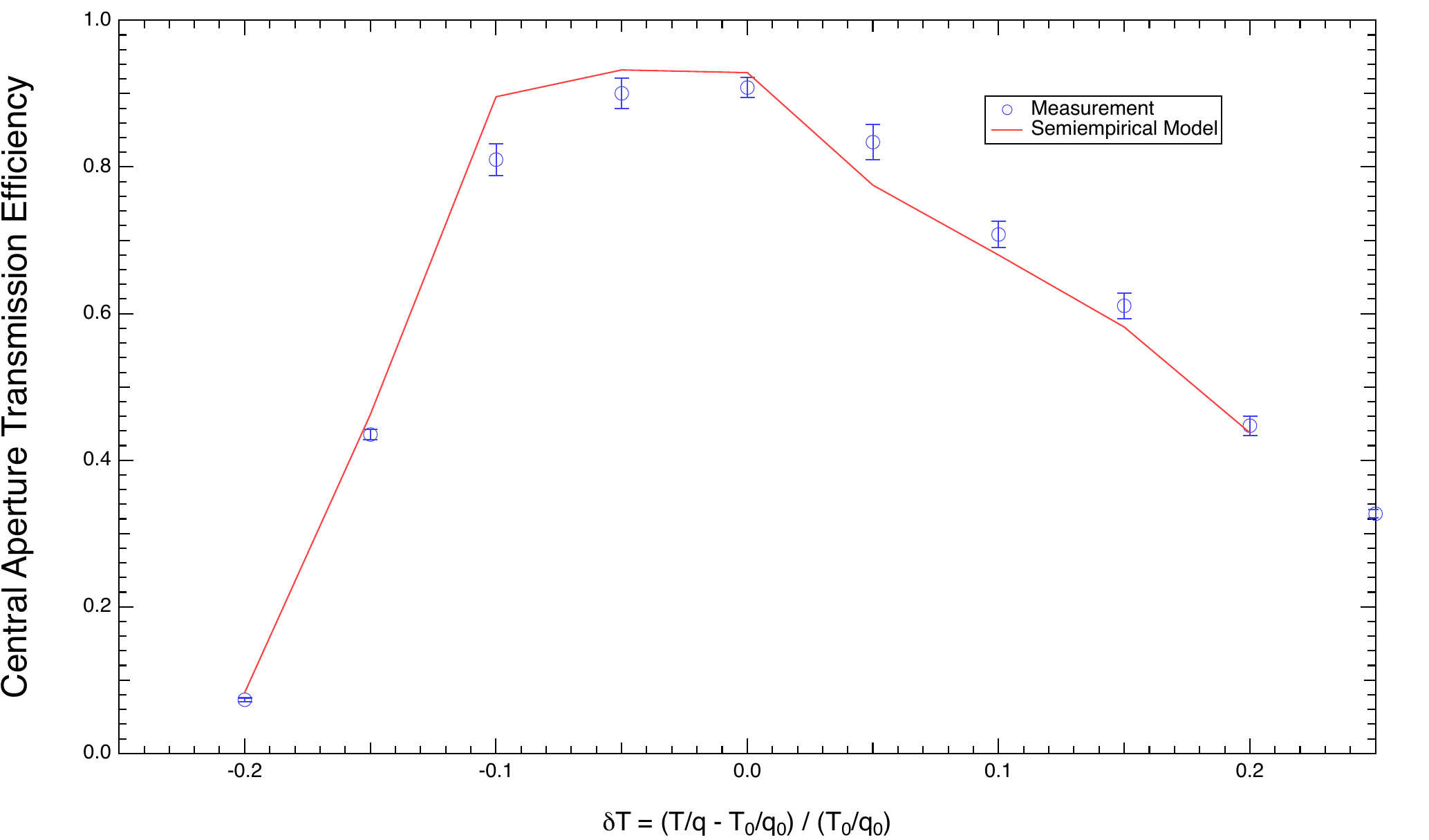}
\caption{Measured mean transmission efficiency as a function of $\delta T$ for the Central angular aperture described in Table \ref{table:apertures} with statistical errors. The solid curve shows the transmission efficiency calculated with the best-fitting empirical model calculations at 9 measured $\delta T$ settings.} \label{fig:central_semi}
 \end{figure}

\begin{figure}
 \includegraphics[width=0.75\textwidth]{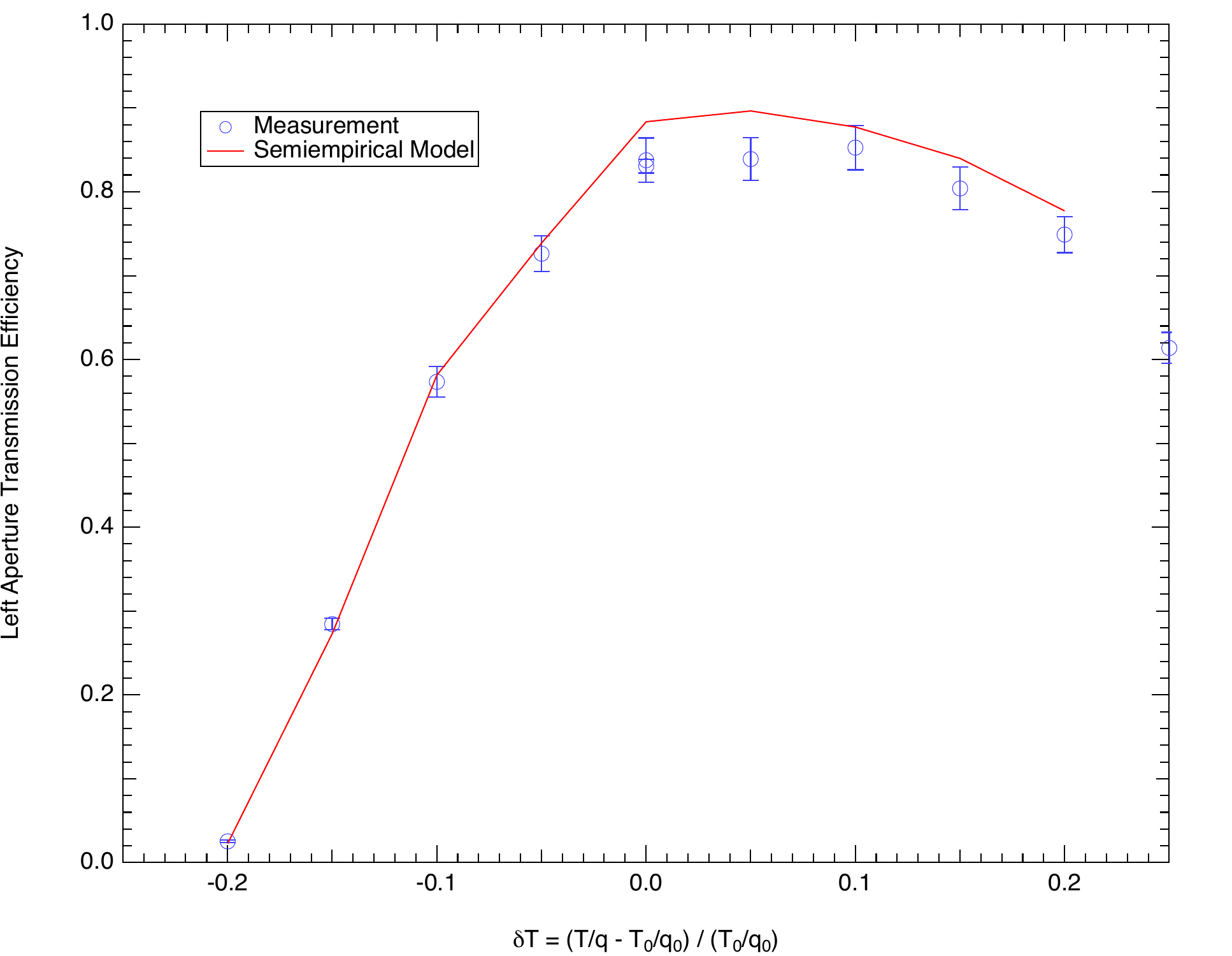}
\caption{Measured mean transmission efficiency as a function of $\delta T$ for the Left angular aperture described in Table \ref{table:apertures} with statistical errors. The solid curve shows the transmission efficiency calculated with the best-fitting empirical model calculations at 9 measured $\delta T$ settings.} \label{fig:left_semi}
 \end{figure}

\begin{figure}
 \includegraphics[width=0.75\textwidth]{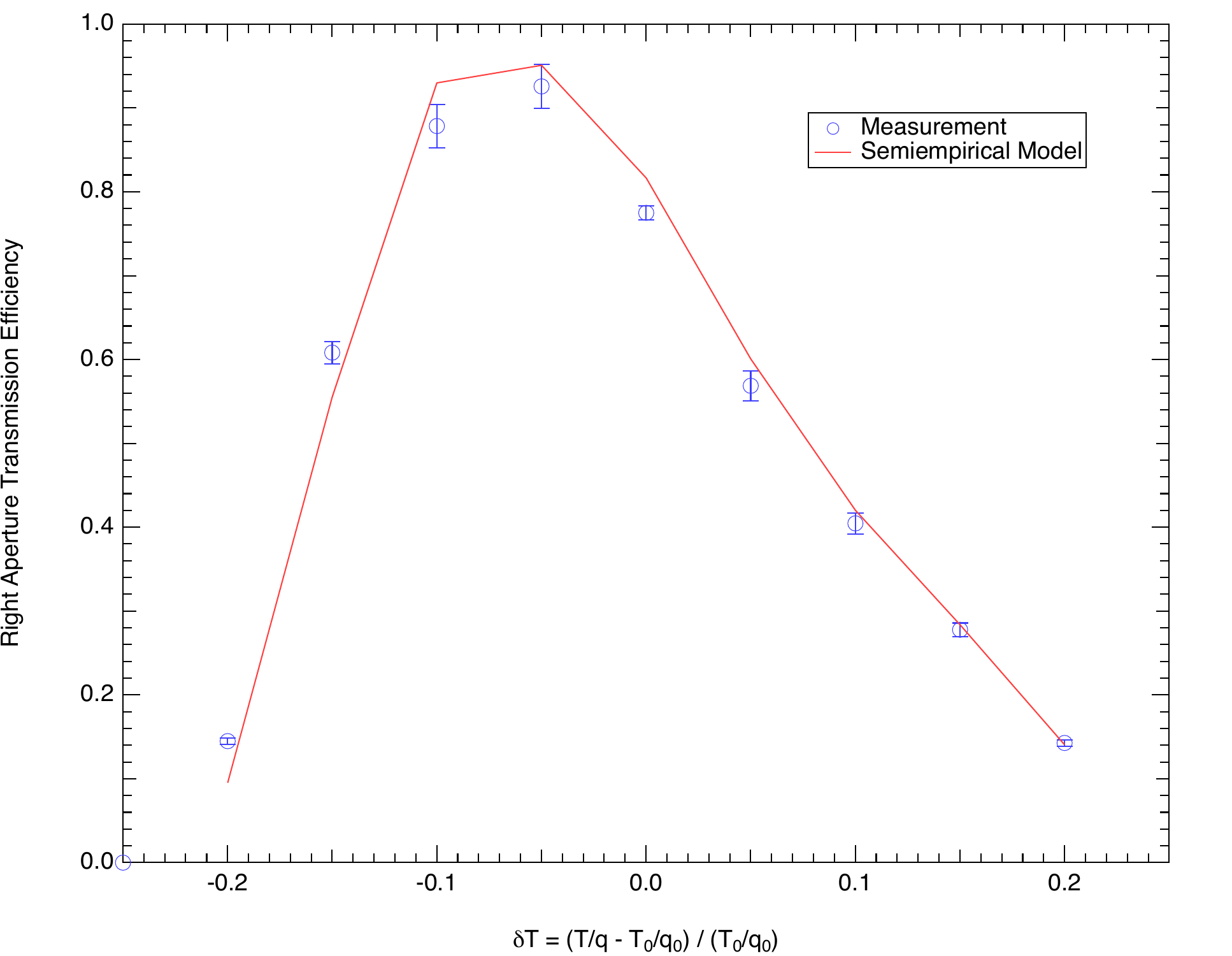}
\caption{Measured mean transmission efficiency as a function of $\delta T$ for the Right angular aperture described in Table \ref{table:apertures} with statistical errors. The solid curve shows the transmission efficiency calculated with the best-fitting empirical model calculations at 9 measured $\delta T$ settings.} \label{fig:right_semi}
 \end{figure}

\begin{figure}
 \includegraphics[width=0.75\textwidth]{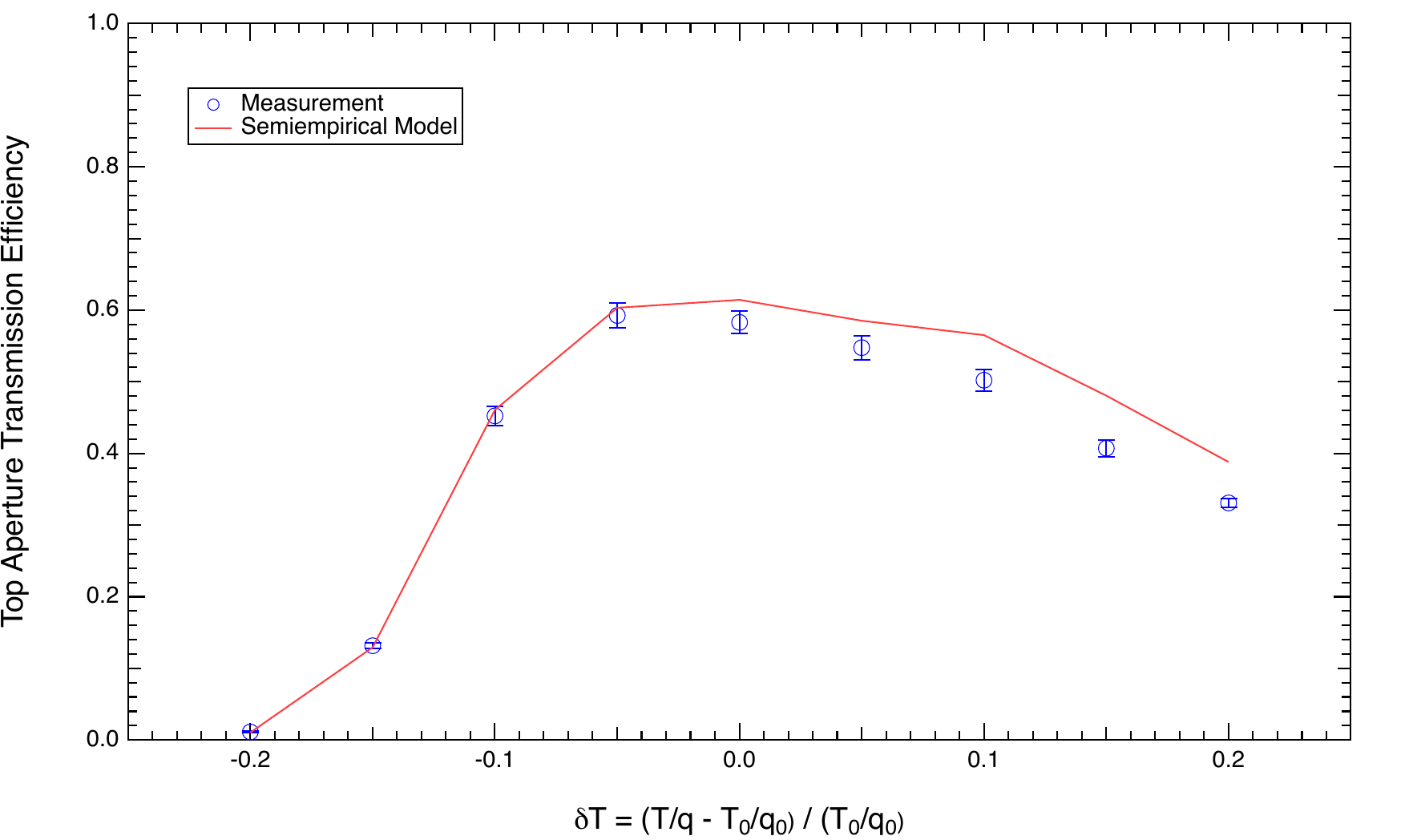}
\caption{Measured mean transmission efficiency as a function of $\delta T$ for the Top angular aperture described in Table \ref{table:apertures} with statistical errors. The solid curve shows the transmission efficiency calculated with the best-fitting empirical model calculations at 9 measured $\delta T$ settings.} \label{fig:top_semi}
 \end{figure}

\begin{figure}
 \includegraphics[width=0.75\textwidth]{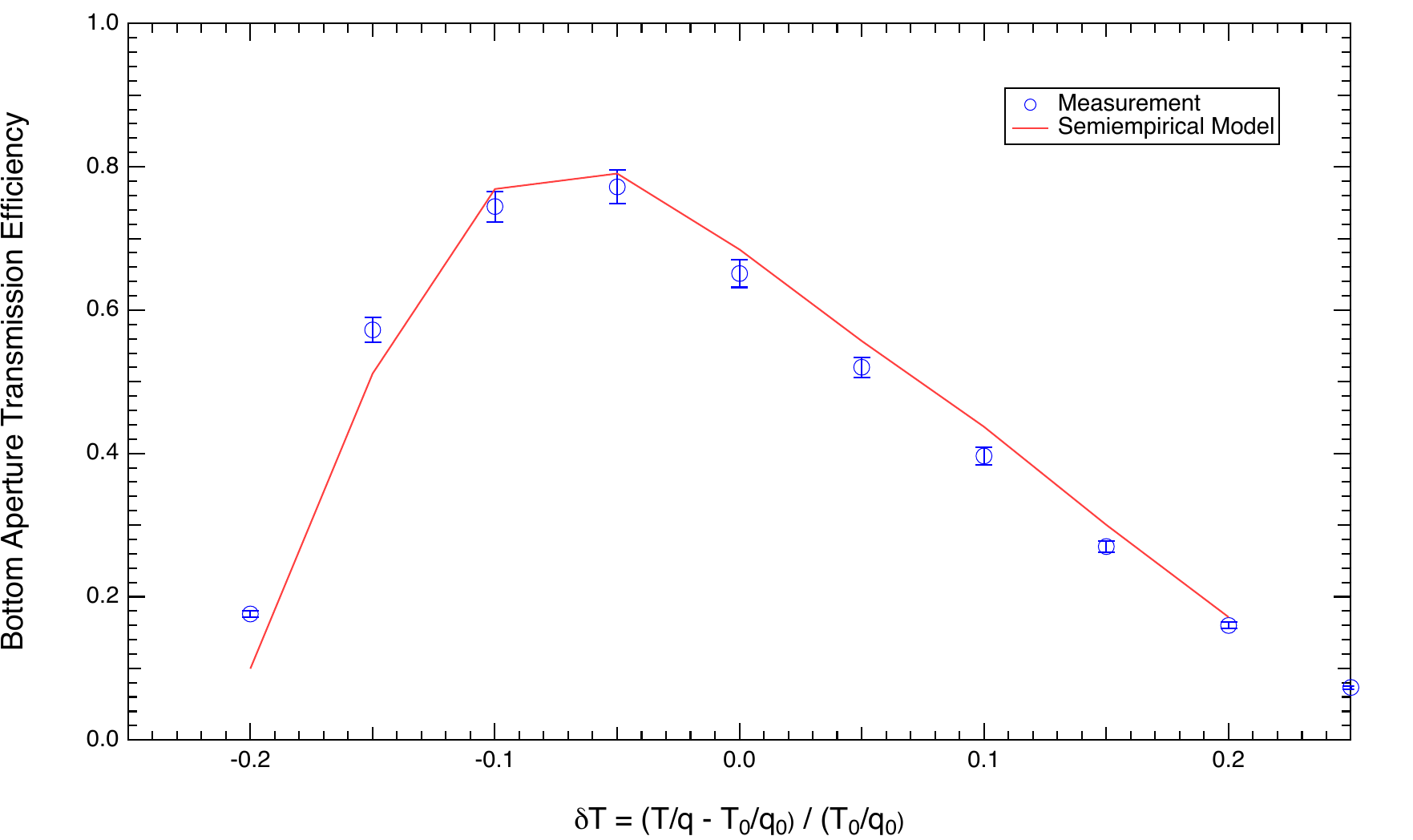}
\caption{Measured mean transmission efficiency as a function of $\delta T$ for the Bottom angular aperture described in Table \ref{table:apertures} with statistical errors. The solid curve shows the transmission efficiency calculated with the best-fitting empirical model calculations at 9 measured $\delta T$ settings.} \label{fig:bottom_semi}
 \end{figure}

The mean transmission efficiencies calculated according to the best fitting empirical model at each of 9 measured $\delta T$ settings for the 6 angular apertures are shown along with the measurements in Figures \ref{fig:full_semi}, \ref{fig:central_semi}, \ref{fig:left_semi}, \ref{fig:right_semi}, \ref{fig:top_semi}, and \ref{fig:bottom_semi}. No single model provided the best fit to the set of 6 measurements at every $\delta T$ setting; the offset piecewise Gaussian was the best model for $\delta T>0$, while the piecewise Gaussian was preferred for the $\delta T \leq 0$ measurements except for the two smallest $\delta T$ settings, which required more parameters to obtain acceptable fits. The difference in the quality of the Gaussian and offset Gaussian model fits of the $\delta T = 0$ measurements is insignificant. The parameters of the best fitting models for 9 $\delta T$ settings within the interval $[-0.2,0.2]$ are shown in Table \ref{table:parameters}; their $\chi^2$ values for the 6 measurements appear in Table \ref{table:uncertainties}.

\begin{table}[h!]
\centering
\begin{tabular}{c c c c c c c c}
	\toprule[1pt]\midrule[0.3pt]
	\noalign{\smallskip}
	$\delta T$ & Gaussian Model & $\sigma_ L$& $\sigma_ R$ & $\sigma_ B$ & $\sigma_T$ & $\theta_0$ & N\\ \midrule
	-0.2 & Normalized Offset Piecewise & $1.93^\circ$ & $0.00002^\circ$ & $4.73^\circ$ & $0.60^\circ$ & $2.21^\circ$ & 0.204\\
	-0.15 & Normalized Offset Piecewise & $2.88^\circ$ & $5.02^\circ$ & $7.89^\circ$ & $0.92^\circ$ & $1.71^\circ$ & 0.641\\
	-0.1 & Piecewise & $1.87^\circ$ & $9.55^\circ$ & $2.77^\circ$ & $1.43^\circ$ &  & \\
	-0.05 & Piecewise & $2.64^\circ$ & $32.94^\circ$ & $2.84^\circ$ & $1.83^\circ$ &  & \\
	0 & Piecewise & $5.36^\circ$ & $3.57^\circ$ & $2.15^\circ$ & $1.86^\circ$ &  & \\
	0.05 & Offset Piecewise & & $5.72^\circ$ & $2.12^\circ$ & $2.29^\circ$ & $-3.68^\circ$ & \\
	0.1 & Offset Piecewise & & $3.93^\circ$ & $1.84^\circ$ & $2.79^\circ$ & $-3.25^\circ$ & \\
	0.15 & Offset Piecewise & & $3.12^\circ$ & $1.45^\circ$ & $2.65^\circ$ & $-3.07^\circ$ & \\
	0.2 & Offset Piecewise & & $2.43^\circ$ & $1.17^\circ$ & $3.02^\circ$ & $-3.03^\circ$ & \\

	\noalign{\smallskip} 
	\midrule[0.3pt]\bottomrule[1pt]
\end{tabular}
\caption{Fitted parameters of the piecewise Gaussian model that best fits the mean transmission efficiencies measured with the 6 angular apertures at each specified $\delta T$ setting.}
\label{table:parameters}
\end{table}

\begin{table}[h!]
\centering
\begin{tabular}{c c c c c}
	\toprule[1pt]\midrule[0.3pt]
	\noalign{\smallskip}
	$\delta T$ & Gaussian Model & $\chi^2$ & Relative Measurement Error & Relative Modelling Error\\ \midrule
	-0.2 & Normalized Offset Piecewise & 31.6 & 0.042 & 1.000\\
	-0.15 & Normalized Offset Piecewise & 3.03 & 0.065 & 0.169\\
	-0.1 & Piecewise & 1.50 & 0.049 & 0.151\\
	-0.05 & Piecewise & 0.32 & 0.042 & 0.053\\
	0 & Piecewise & 1.58 & 0.043 & 0.050\\
	0.05 & Offset Piecewise & 2.56 & 0.048 & 0.075\\
	0.1 & Offset Piecewise & 4.92 & 0.053 & 0.131\\
	0.15 & Offset Piecewise & 8.46 & 0.043 & 0.193\\
	0.2 & Offset Piecewise & 6.67 & 0.054 & 0.321\\

	\noalign{\smallskip} 
	\midrule[0.3pt]\bottomrule[1pt]
\end{tabular}
\caption{Total $\chi^2$ value and the relative measurement and modelling uncertainties associated with the piecewise Gaussian model that best fits the mean transmission efficiencies measured with the 6 angular apertures at each specified $\delta T$ setting. Due to the poor quality of the fit of the $\delta T=-0.2$ data, we assign a 100\% relative modelling uncertainty to that model.}
\label{table:uncertainties}
\end{table}

\begin{figure}
\begin{subfigure}{0.5\textwidth}
 \includegraphics[width=\linewidth, height=0.797\linewidth]{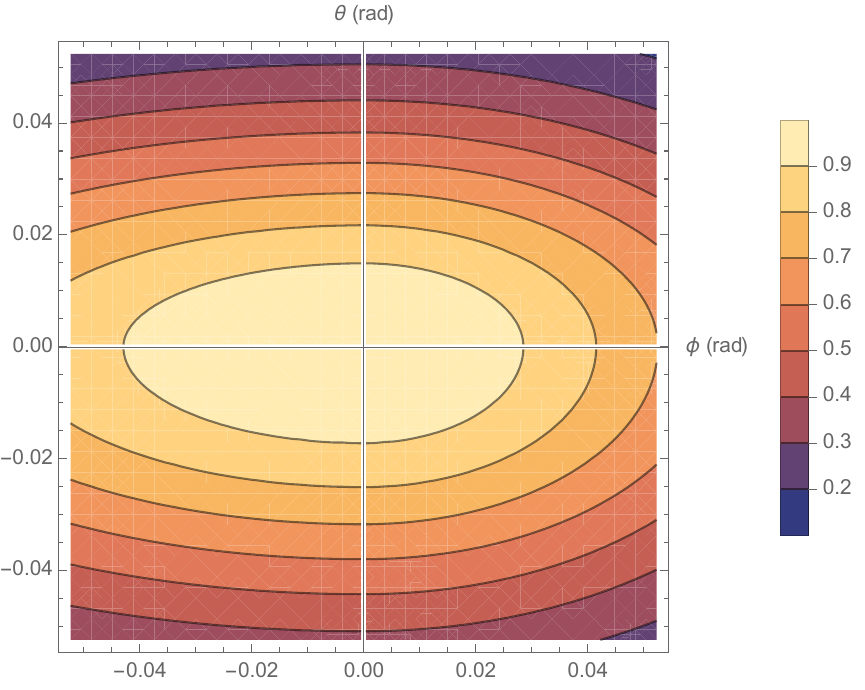}
\caption{Piecewise Gaussian}
 \label{contour_pg_0}
\end{subfigure}
\begin{subfigure}{0.5\textwidth}
 \includegraphics[width=\linewidth, height=0.797\linewidth]{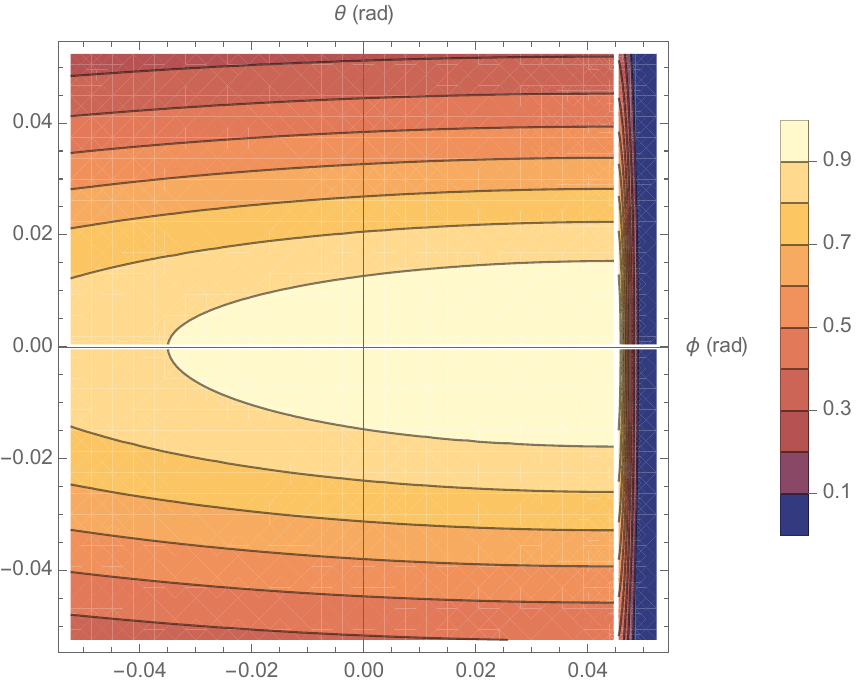}
\caption{Offset Piecewise Gaussian}
 \label{contour_og_0}
\end{subfigure}
\caption{Contour plot of the EMMA transmission efficiency for $\delta T = 0$ calculated according to models whose parameters best fit the 6 angular aperture measurements as a function of $\theta$ and $\phi$, plotted over the Full angular aperture.}
\label{contour_0}
 \end{figure}

\begin{figure}
\begin{subfigure}{0.5\textwidth}
 \includegraphics[width=\linewidth, height=0.797\linewidth]{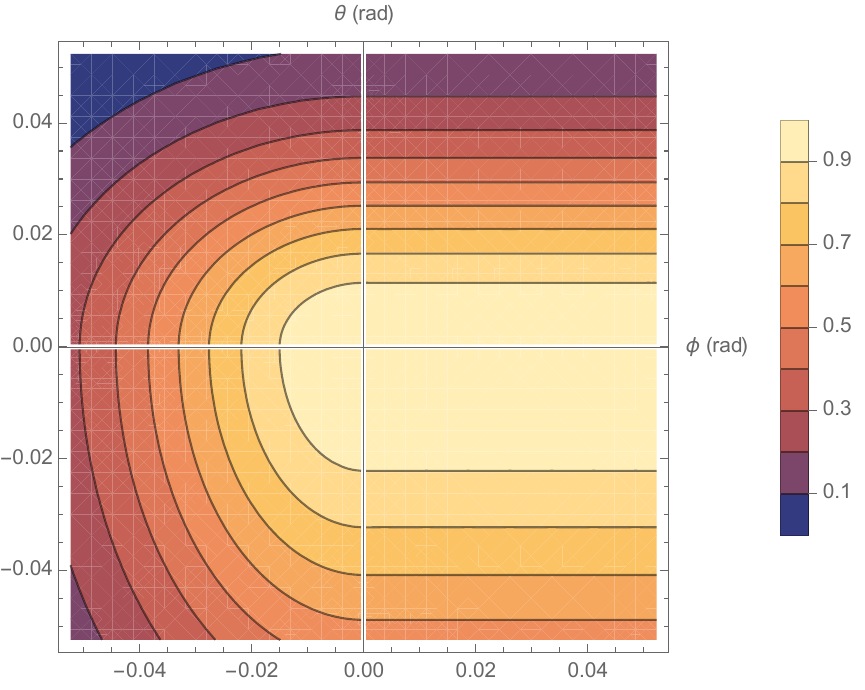}
\caption{Piecewise Gaussian}
 \label{contour_pg_-10}
\end{subfigure}
\begin{subfigure}{0.5\textwidth}
 \includegraphics[width=\linewidth, height=0.797\linewidth]{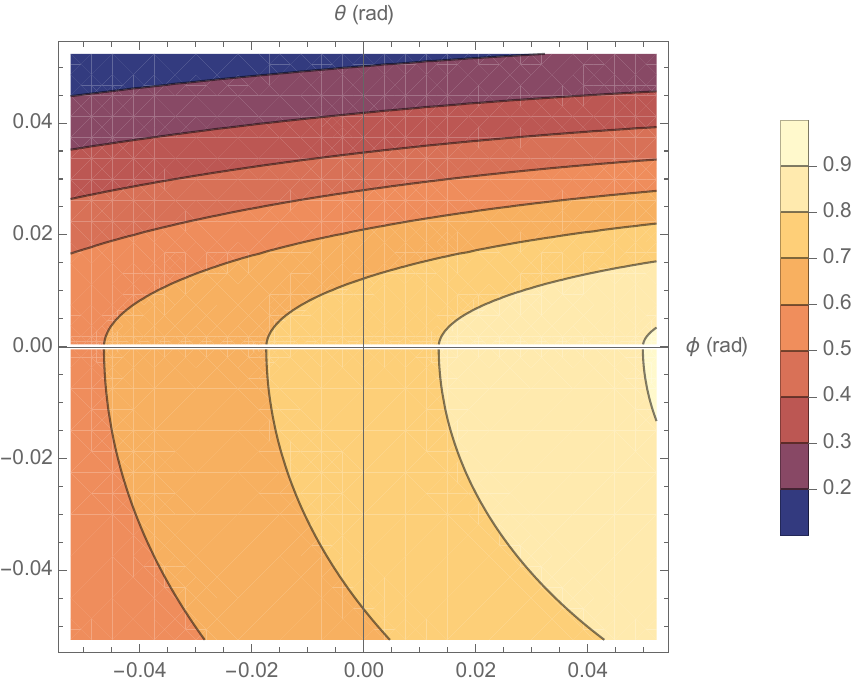}
\caption{Offset Piecewise Gaussian}
 \label{contour_og_-10}
\end{subfigure}
\caption{Contour plot of the EMMA transmission efficiency for $\delta T = -0.1$ calculated according to models whose parameters best fit the 6 angular aperture measurements as a function of $\theta$ and $\phi$, plotted over the Full angular aperture.}
\label{contour_-10}
 \end{figure}

\begin{figure}
\begin{subfigure}{0.5\textwidth}
 \includegraphics[width=\linewidth, height=0.797\linewidth]{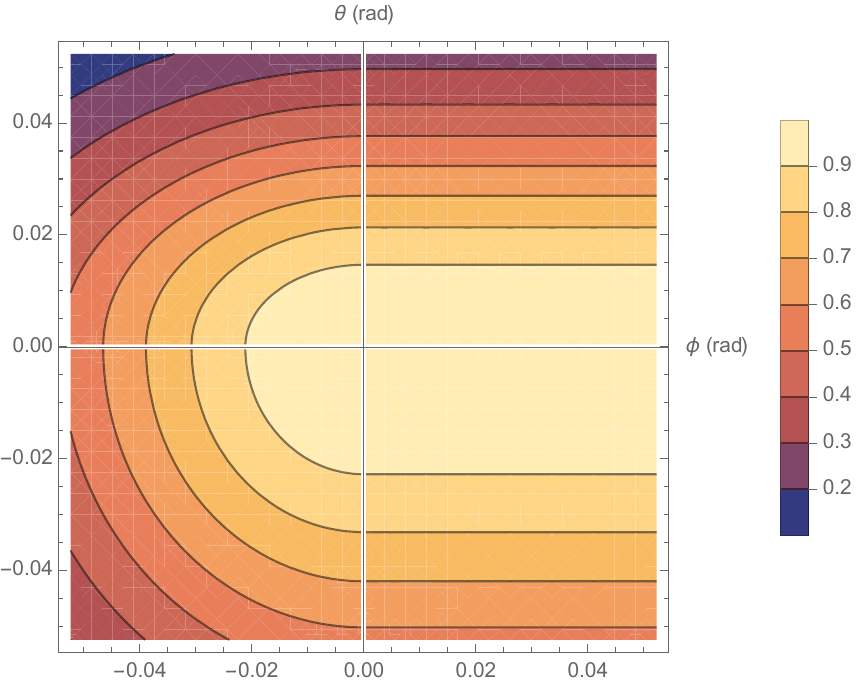}
\caption{Piecewise Gaussian}
 \label{contour_pg_-5}
\end{subfigure}
\begin{subfigure}{0.5\textwidth}
 \includegraphics[width=\linewidth, height=0.797\linewidth]{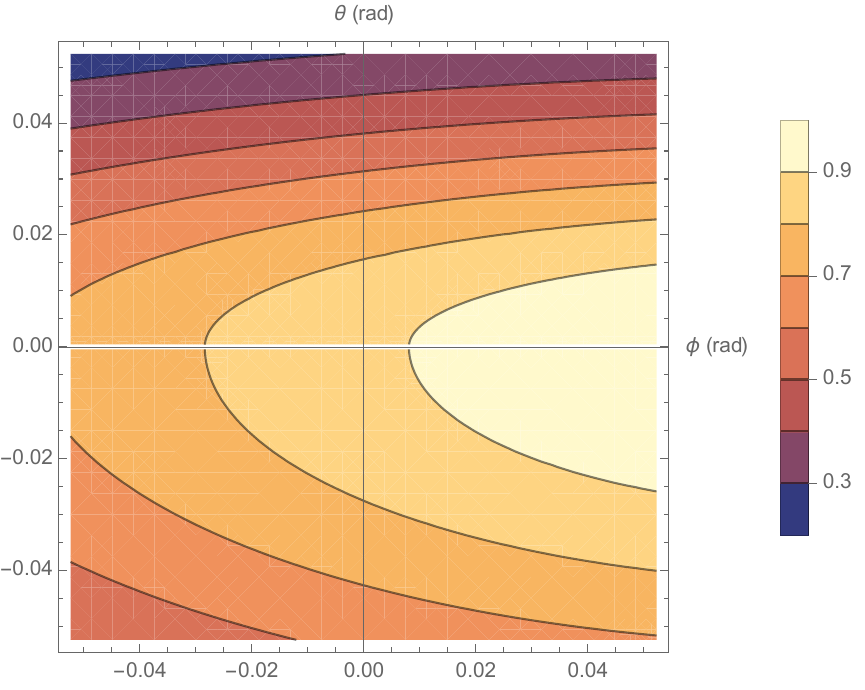}
\caption{Offset Piecewise Gaussian}
 \label{contour_og_-5}
\end{subfigure}
\caption{Contour plot of the EMMA transmission efficiency for $\delta T = -0.05$ calculated according to models whose parameters best fit the 6 angular aperture measurements as a function of $\theta$ and $\phi$, plotted over the Full angular aperture.}
\label{contour_-5}
 \end{figure}

\begin{figure}
\begin{subfigure}{0.5\textwidth}
 \includegraphics[width=\linewidth, height=0.797\linewidth]{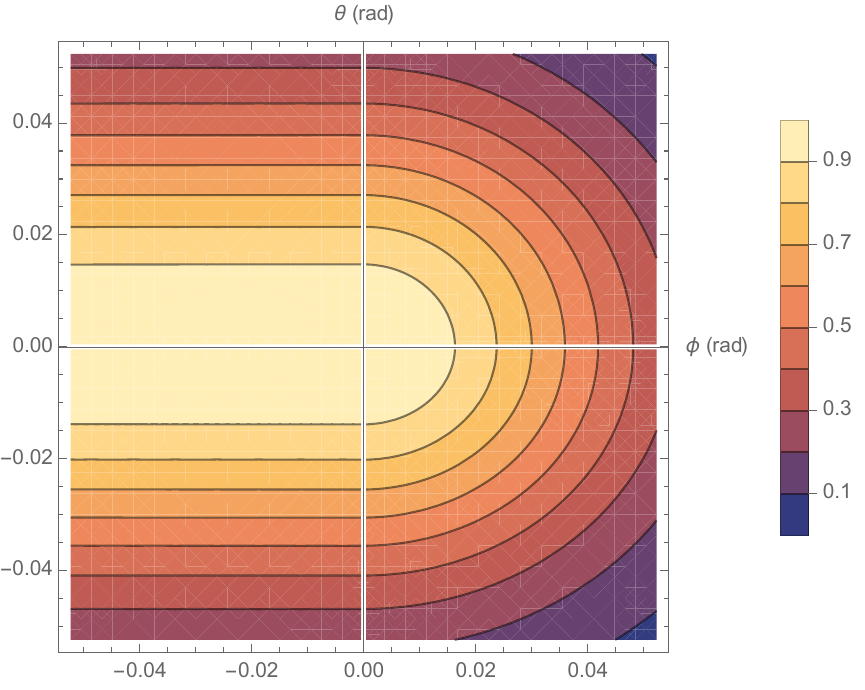}
\caption{Piecewise Gaussian}
 \label{contour_pg_5}
\end{subfigure}
\begin{subfigure}{0.5\textwidth}
 \includegraphics[width=\linewidth, height=0.797\linewidth]{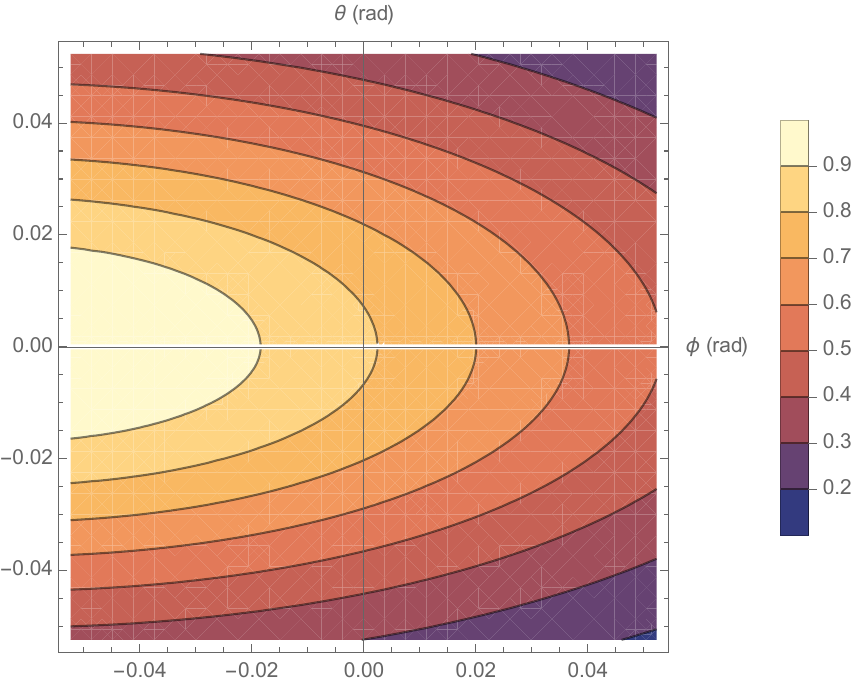}
\caption{Offset Piecewise Gaussian}
 \label{contour_og_5}
\end{subfigure}
\caption{Contour plot of the EMMA transmission efficiency for $\delta T = 0.05$ calculated according to models whose parameters best fit the 6 angular aperture measurements as a function of $\theta$ and $\phi$, plotted over the Full angular aperture.}
\label{contour_5}
 \end{figure}

\begin{figure}
\begin{subfigure}{0.5\textwidth}
 \includegraphics[width=\linewidth, height=0.793\linewidth]{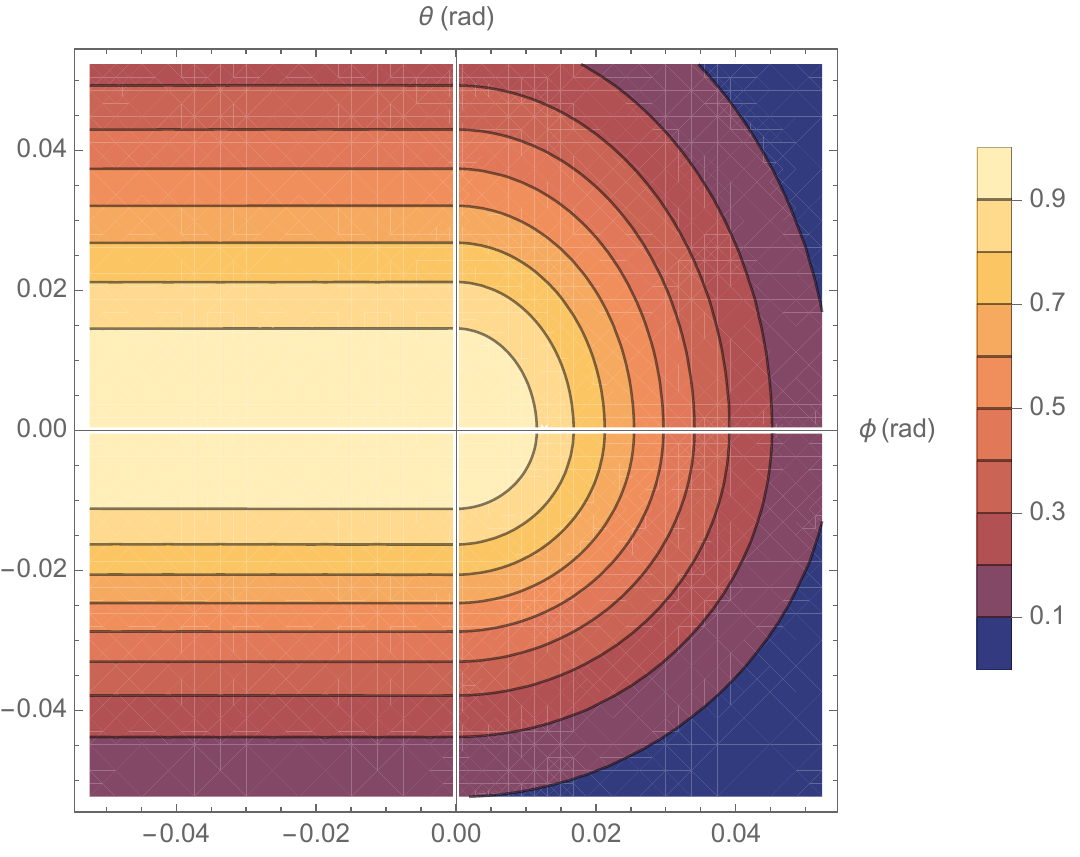}
\caption{Piecewise Gaussian}
 \label{contour_pg_10}
\end{subfigure}
\begin{subfigure}{0.5\textwidth}
 \includegraphics[width=\linewidth, height=0.793\linewidth]{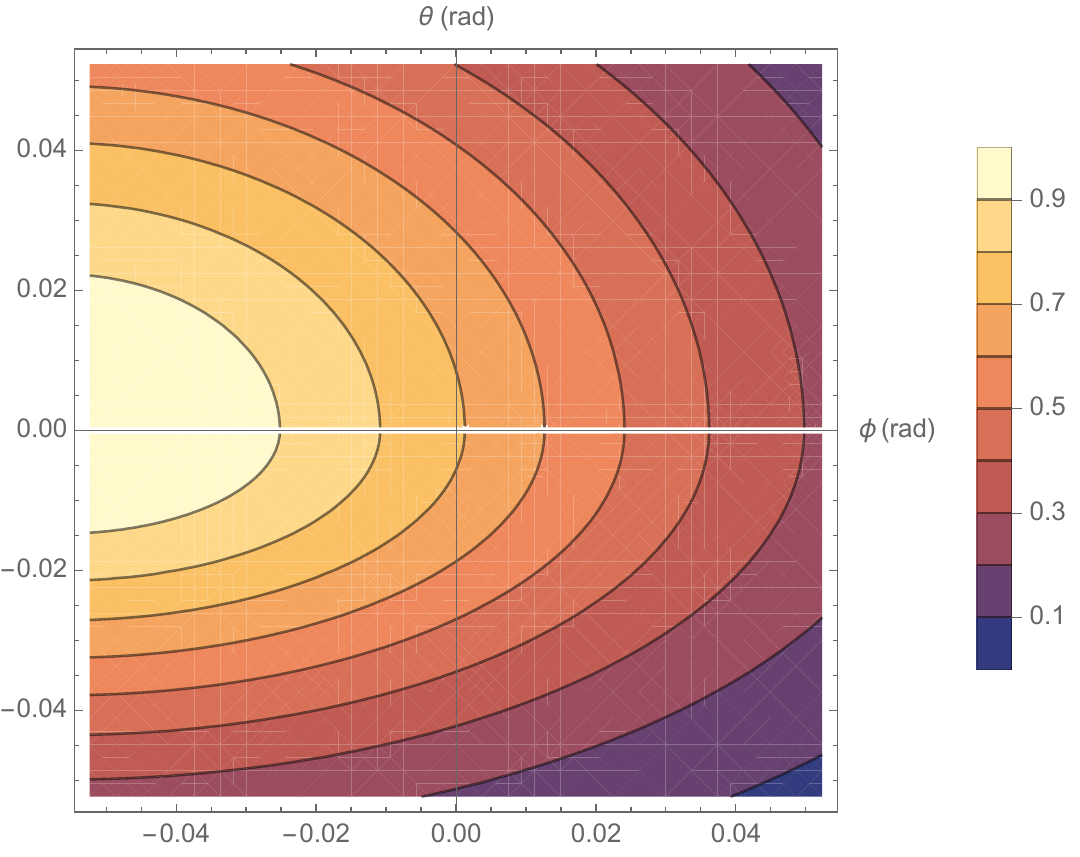}
\caption{Offset Piecewise Gaussian}
 \label{contour_og_10}
\end{subfigure}
\caption{Contour plot of the EMMA transmission efficiency for $\delta T = 0.1$ calculated according to models whose parameters best fit the 6 angular aperture measurements as a function of $\theta$ and $\phi$, plotted over the Full angular aperture.}
\label{contour_10}
 \end{figure}

Figure \ref{contour_0} depicts contour plots of the transmission efficiency for the $\delta T = 0$ case, calculated according to a piecewise Gaussian function and an offset piecewise Gaussian function, both of whose parameters best fit the 6 angular aperture measurements for $\delta T = 0$, as a function of $\theta$ and $\phi$, plotted over the Full angular aperture. The equivalent plots for the $\delta T = -0.1$, $\delta T = -0.05$, $\delta T = 0.05$, and $\delta T = 0.1$ cases are shown in Figures \ref{contour_-10}, \ref{contour_-5}, \ref{contour_5}, and \ref{contour_10}, respectively. 

No attempt to systematically measure or model the transmission efficiency at angles beyond those subtended by the Full angular aperture was made. Downstream of the angular aperture, the target chamber has an exit aperture that describes a cone of half angle $4.2^\circ$. Hence the transmission efficiency at polar scattering angles $\Theta > 4.2^\circ$ vanishes. In the absence of additional information, the transmission efficiency at angles beyond the extent of the Full angular aperture but within the circular exit aperture is calculated as the average of the value derived from the best fitting model within the Full angular aperture and 0, since the former is a reliable upper limit and the latter a certain lower limit of the transmission efficiency at these angles.

Given the models that adequately describe the mean transmission efficiency at 9 $\delta T$ settings between -0.2 and 0.2, to find the transmission efficiency at arbitrary $\theta$, $\phi$, and $\delta T$, we interpolate linearly between the models at the two $\delta T$ settings that are closest to the $\delta T$ value of interest. Extrapolation beyond the modelled range of $\delta T \in [-0.2,0.2]$ is not required because no EMMA experiments producing recoils with such large $\delta T$ values require absolute cross section determinations. In fact, no cross section measurement with recoils having $|\delta T|>0.1$ has been performed or is planned.

When estimating the total uncertainty of the transmission efficiency for any particular combination of $\theta$, $\phi$, and $\delta T$, statistical and systematic contributions must be considered. We compute the relative error in the transmission efficiency $\epsilon(\delta T,\theta, \phi)$ by linearly interpolating the relative errors in $\epsilon(\theta, \phi)$ at the two measured $\delta T$ settings that are closest to the $\delta T$ of interest. The relative uncertainty in $\epsilon(\theta, \phi)$ for a measured value of $\delta T$ is estimated as a quadratic sum of 3 terms.

\begin{figure}
 \includegraphics[width=0.75\textwidth]{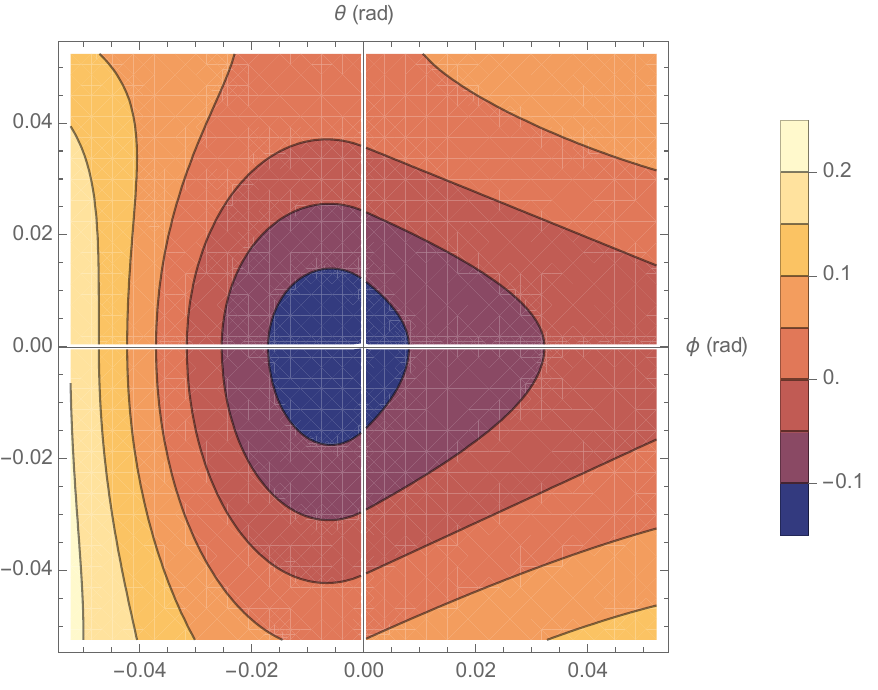}
\caption{Contour plot of the difference between the transmission efficiencies for $\delta T = -0.05$ calculated according to the best-fitting offset and piecewise Gaussian models as a function of $\theta$ and $\phi$, plotted over the Full angular aperture.} \label{fig:diff}
 \end{figure}

The first is the relative uncertainty of the mean transmission efficiency measurements with the Full angular aperture, which includes both statistical and systematic contributions and ranges from 4--6\% at the various $\delta T$ settings. The second term is the relative systematic modelling uncertainty. We estimate its size using the difference between the best fitting models for each value of $\delta T$, $\theta$, and $\phi$. For example, Fig.\ \ref{fig:diff} shows the difference between the transmission efficiency for $\delta T = -0.05$ calculated according to the best-fitting offset and piecewise Gaussian models as a function of $\theta$ and $\phi$, plotted over the Full angular aperture; the average difference between the transmission efficiencies calculated with the two $\delta T = -0.05$ models is 0.012 and the root mean square difference is 0.074 over the angular range of $\pm3^\circ$ by $\pm3^\circ$. The relative systematic modelling uncertainty is taken to be 1/3 of the absolute value of the relative difference between the best fitting models, averaged over the Full angular aperture, and ranges from a minimum of 5\% at  $\delta T=0$ to a local maximum of 32\% at $\delta T=0.2$; an exception is $\delta T=-0.2$, where no good fit was obtained. In this case we assigned a 100\% relative modelling uncertainty. Table \ref{table:uncertainties} lists the relative measurement and modelling uncertainties for the adopted transmission efficiency models at each of the specified measured $\delta T$ values.

The final term in the quadratic sum is the relative error due to extrapolation beyond the Full angular aperture and accordingly is only finite for trajectories with $|\theta|>3^\circ$ or $|\phi|>3^\circ$ (0.052~rad). The absolute uncertainty in this quantity is taken to be 1/3 of the difference between the extrapolated best fitting model and 0. Since the transmission efficiency is assumed to be the average of the two, the relative uncertainty due to extrapolation beyond the Full angular aperture is 2/3.
 
Having modelled the transmission efficiency of monoisotopic ions as a function of their kinetic energy, charge, and scattering angles, $\epsilon(\delta T,\theta, \phi)$, we are now in a position to evaluate the mean transmission efficiency for the recoils of any particular measurement. This is achieved in two steps. First, the momenta of the recoils from the reaction of interest must be simulated, properly taking into account the beam emittance, target composition and thickness, etc. In practice, we typically utilize GEANT4 or LISE$^{++}_{cute}$ \cite{tarasov23} for these simulations and generate a file containing the $\delta T$, $\theta$, and $\phi$ of the recoils. This file serves as the input for an independent computer code that calculates the transmission efficiency and associated uncertainty of each simulated recoil and computes the mean transmission efficiency and uncertainty for the whole distribution.

\section{Summary} \label{sec:concl}

In summary, the mean transmission efficiencies of the EMMA recoil mass spectrometer at TRIUMF were measured as a function of kinetic energy/charge over various ranges of the horizontal and vertical projections of the scattering angle. All measurements were performed using a monoenergetic $\alpha$ source installed at the target location, with 6 angular apertures positioned at the entrance of the spectrometer, taking advantage of the scaling properties of the trajectories of ions moving through electromagnetic fields. The spectrometer was set for measurements at 17 different kinetic energy/charge values. After comparing the results of measurements at 9 regularly spaced kinetic energy/charge deviations between -20\% and 20\% of the reference trajectory with Monte Carlo simulations and finding discrepancies, we adopted an empirical approach, describing the transmission efficiency as a function of the two angles at each of these 9 settings. Linear interpolation between the best fitting empirical piecewise Gaussian models allows for accurate estimates of the transmission efficiency of the spectrometer and its systematic uncertainty over the entire range of energies and angles of interest for absolute cross section measurements.

\section*{Acknowledgements}

This work is dedicated to the memory of Birger B.\ Back. The authors acknowledge the generous support of the Natural Sciences and Engineering Research Council of Canada. TRIUMF receives federal funding via a contribution agreement through the National Research Council of Canada. MW gratefully acknowledges support from the UK Science and Technologies Facilities Council (STFC). We thank Peter Gumplinger for help with GEANT4 and Sara Moll\'o and Laura Pedro-Botet for Python programming.


\bibliographystyle{unsrt}
\bibliography{emma.bib}

\begin{thebibliography}{10}

\bibitem{hackman14}
G.~{Hackman} and C.~E. {Svensson}.
\newblock {\em Hyperfine Interactions}, 225:241, 2014.

\bibitem{davids19}
B.~{Davids}, M.~{Williams}, N.E. {Esker}, M.~{Alcorta}, et~al.
\newblock {\em Nucl. Instrum. Meth. Phys. Res. A}, 930:191, 2019.

\bibitem{lotay21}
G.~{Lotay}, S.~A. {Gillespie}, M.~{Williams}, T.~{Rauscher}, et~al.
\newblock {\em Phys. Rev. Lett.}, 127:112701, 2021.

\bibitem{williams23}
M.~Williams, B.~Davids, G.~Lotay, N.~{Nishimura}, T.~{Rauscher}, S.~A.
  {Gillespie}, M.~{Alcorta}, et~al.
\newblock {\em Phys. Rev. C}, 107:035803, 2023.

\bibitem{williams25}
M.~{Williams}, C.~{Angus}, A.~M. {Laird}, B.~{Davids}, C.~Aa. {Diget},
  A.~{Fernandez}, E.~J. {Williams}, et~al.
\newblock {\em Phys. Rev. Lett.}, 134:112701, 2025.

\bibitem{wollnik87a}
H.~Wollnik.
\newblock {\em Optics of Charged Particles}.
\newblock Academic Press, Inc., 1987.

\bibitem{back96}
B.~B. Back et~al.
\newblock {\em Nucl. Instrum. Methods Phys. Res.}, A 379:206, 1996.

\bibitem{akovali98}
Y.~A. Akovali.
\newblock {\em Nuclear Data Sheets}, 84:1, 1998.

\bibitem{nica14}
N.~Nica.
\newblock {\em Nuclear Data Sheets}, 117:1, 2014.

\bibitem{geant4_ref}
S.~Agostinelli et~al.
\newblock {\em Nucl. Instrum. Meth. Phys. Res. A}, 506:250, 2003.

\bibitem{davids05}
B.~{Davids} and C.~N. {Davids}.
\newblock {\em Nucl. Instrum. Meth. Phys. Res. A}, 544:565, 2005.

\bibitem{wollnik87}
H.~Wollnik, J.~Brezina, and M.~Berz.
\newblock {\em Nucl. Instrum. Methods Phys. Res.}, A258:408, 1987.

\bibitem{james75}
F.~{James} and M.~{Roos}.
\newblock {\em Computer Physics Communications}, 10:343--367, 1975.

\bibitem{tarasov23}
O.~B. Tarasov, D.~Bazin, et~al.
\newblock {\em Nucl. Instrum. Meth. Phys. Res. B}, 541:4, 2023.

\end{thebibliography}






\end{document}